\documentclass[%
 reprint,
showpacs,
nofootinbib,
 amsmath,amssymb,
 aps,
floatfix,
]{revtex4-1}

\usepackage[dvipsnames]{xcolor}
\usepackage{soul}
\usepackage{graphicx}
\usepackage{dcolumn}
\usepackage{bm}
\usepackage{verbatim}
\usepackage[caption=false]{subfig}
\usepackage[colorlinks=true]{hyperref}
\hypersetup{
    colorlinks,
    linkcolor=blue!50!black,          
    citecolor=blue!50!black,        
    urlcolor=blue!50!black           
}

\graphicspath{{NSimages/}}

\begin{document}

\preprint{APS/123-QED}

\title{Neutrino Flavor Evolution in Neutron Star Mergers}

\author{James Y. Tian}
 \email{j2tian@ucsd.edu}
 \author{Amol V. Patwardhan}
 \email{apatward@ucsd.edu}
\author{George M. Fuller}%
 \email{gfuller@ucsd.edu}
\affiliation{%
Department of Physics, University of California, San Diego, La Jolla, California 92093-0319, USA
}%




\date{\today}

\begin{abstract}
We examine the flavor evolution of neutrinos emitted from the disk-like remnant (hereafter called \lq\lq neutrino disk\rq\rq) of a binary neutron star (BNS) merger. We specifically follow the neutrinos emitted from the center of the disk, along the polar axis perpendicular to the equatorial plane. We carried out two-flavor simulations using a variety of different possible initial neutrino luminosities and energy spectra, and for comparison, three-flavor simulations in specific cases. In all simulations, the normal neutrino mass hierarchy was used. The flavor evolution was found to be highly dependent on the initial neutrino luminosities and energy spectra; in particular, we found two broad classes of results depending on the sign of the initial net electron neutrino lepton number (i.e., the number of neutrinos minus the number of antineutrinos). In the antineutrino dominated case, we found that the Matter-Neutrino Resonance (MNR) effect dominates, consistent with previous results, whereas in the neutrino dominated case, a bipolar spectral swap develops. The neutrino dominated conditions required for this latter result have been realized, e.g, in a BNS merger simulation that employs the \lq\lq DD2\rq\rq\ equation of state for neutron star matter~\cite{2016PhRvD..93d4019F}. For this case, in addition to the swap at low energies, a collective Mikheyev-Smirnov-Wolfenstein (MSW) mechanism generates a high-energy electron neutrino tail. The enhanced population of high-energy electron neutrinos in this scenario could have implications for the prospects of $r$-process nucleosynthesis in the material ejected outside the plane of the neutrino disk.
\end{abstract}
\pacs{14.60.Pq, 97.60.Bw, 13.15.+g, 26.30.-k, 26.50.+x}
\maketitle


\section{Introduction}

In this paper we explore neutrino flavor transformation in the binary neutron star (BNS) merger environment, for axially directed neutrinos, and with an emphasis on scenarios with a higher number luminosity of neutrinos over antineutrinos. Depending on the BNS merger rate and on the amount of ejected material, BNS merger events could be a potential candidate site for the origin of nuclei heavier than $^{56}$Fe via the $r$-process~\cite{2016MNRAS.460.3255R, 2016Natur.531..610J, 1999ApJ...525L.121F, 2014MNRAS.439..744R, 2016PhRvD..94l3016F, 1982ApL....22..143S, 2014MNRAS.443.3134P, 2011ApJ...738L..32G, 1977ApJ...213..225L, 1989ApJ...343..254M, 1974ApJ...192L.145L, 2015MNRAS.447..140V, 2016MNRAS.463.2323W, 2015ApJ...813....2M, 1997A&A...319..122R}. These cataclysmic events are accompanied by prodigious fluxes of neutrinos~\cite{2016PhRvD..93d4019F, 2016MNRAS.460.3255R, 2009ApJ...690.1681D, 2014MNRAS.443.3134P, 1997A&A...319..122R}. Flavor-dependent charged current neutrino capture reactions could influence the neutron content of these ejecta~\cite{Qian93}, depending on the material outflow speed and geometry, and on the neutrino luminosities, energy spectra, and emission geometry. Though many of these ingredient quantities have not yet been unambiguously determined by simulations, the importance and urgency of the $r$-process origin problem warrants an exploration of neutrino flavor physics in this environment.

There are roughly three potential sources of $r$-process material in BNS mergers: (1) the \lq\lq tidal tails\rq\rq\ of neutron-rich nuclear matter tidally stripped from the neutron stars during the in-spiral event before the stars touch; (2) the material ejected in the equatorial disk formed when the stars have merged; and (3) the material driven off by either magneto-hydrodynamic (MHD) mechanisms or intense neutrino radiation, in directions outside the equatorial plane (e.g., along the polar axis). Of these, only the material in the first will not be accompanied by significant neutrino and antineutrino radiation exposure.

In order to evaluate the efficacy of BNS mergers in producing the observed $r$-process abundances, it is essential to have good estimates of merger rates over cosmic history, along with the average $r$-process yield per merger. Current BNS merger rate estimates are extremely primitive, as they are based on a very small sample size of observed binary pulsars in the Milky Way~\cite{2003Natur.426..531B, 2003ApJ...584..985K,2001ApJ...556..340K, 1999ApJ...520..696A, 1996MNRAS.283L..37V, 2012LRR....15....8F, 2015ApJ...814...58D, 1991ApJ...380L..17P}. However, recent direct detections of gravitational waves from binary black-hole merger events by the Laser Interferometer Gravitational-wave Observatory (LIGO)~\cite{2016PhRvL.116f1102A,2016PhRvL.116x1103A} have opened up an entirely new channel for exploring the universe. The current estimated upper limit on the BNS merger rate from LIGO is 12,600 $\text{Gpc}^{-3}\text{yr}^{-1}$ (at $90$\% C.L.), based on not having observed the gravitational wave signal from a BNS merger event yet~\cite{2016ApJ...832L..21A}. This limit is consistent with the current binary pulsar based estimates. As LIGO begins to reach towards its ultimate design sensitivity within the next few years, it is hoped that it will enable us to obtain much better estimates, or at least more stringent upper bounds, on the rates of BNS merger events in the present-day universe~\cite{2017ApJ...836..230C}.

While we wait for LIGO to give us a better observational handle on the merger rate, we can examine the other aspect of the problem by looking at the $r$-process yields of individual BNS merger events. Previous work has attempted to quantify the amount of $r$-process yields in the ejecta of neutron star mergers~\cite{2016MNRAS.460.3255R, 1999ApJ...525L.121F, 2014MNRAS.439..744R, 2016PhRvD..94l3016F, 2014MNRAS.443.3134P, 2011ApJ...738L..32G, 2016MNRAS.463.2323W,2015ApJ...813....2M, 1997A&A...319..122R}. However, the neutrino physics that goes into these simulations is primitive at best, and in particular does not include a treatment of flavor conversion in these environments. Neutrinos are emitted with very high luminosities (on the order of $10^{53} \, \text{erg}\,\text{s}^{-1}$) from the disk-like (temporary) remnant of the neutron star merger. As long as the weak interactions are not fully decoupled, these neutrinos, via charged-current capture on nucleons, can determine the electron fraction ($Y_e$) of the material they interact with. This, in turn, is a major factor in evaluating the feasibility of these events as $r$-process producers. The neutrinos affect the electron fraction via the following neutrino and antineutrino capture reactions:
\begin{subequations}
\begin{align}
\nu_e+n\rightleftharpoons p + e^-, \label{eqn:ntop} \\ 
\bar{\nu}_e+p\rightleftharpoons n + e^+.\label{eqn:pton}
\end{align} 
\label{eqn:reactions}
\end{subequations}

Since the neutrino charged-current capture processes are flavor-asymmetric at typical energies ($\sim$ 10 MeV), i.e., only (anti-)neutrinos in the electron flavor state can participate, it is essential that we know the flavor histories of these neutrinos as they stream out of the merger site. A large change in the flavor content of neutrinos or even just in their energy spectra (since the capture rates are energy dependent) could have a correspondingly large effect on the electron fraction and therefore on the efficacy of the $r$-process in these environments. Therefore, detailed analysis of neutrino flavor evolution is necessary in order to better understand the potential these environments have for being the main sites of the $r$-process.

Indeed, recent explorations of neutrino flavor evolution have found collective phenomena in certain regions of BNS merger outflow~\cite{2016arXiv160705938F, 2014arXiv1403.5797M, 2016arXiv160705938F, 2016arXiv160704671Z, 2016PhRvD..93d5021M, 2016arXiv161101862C, 2017arXiv170106580W}. Most of these consider initial conditions that exhibit an overall anti-neutrino number dominance over neutrinos. The various types of flavor transformation phenomena found in these calculations include symmetric and standard Matter-Neutrino-Resonances (MNR)~\cite{2012PhRvD..86h5015M, 2014arXiv1403.5797M, 2016arXiv160705938F, 2016arXiv160704671Z, 2016PhRvD..93d5021M, 2016PhLB..752...89W,  2016PhRvD..93j5044V, 2016arXiv160504903S}, fast pairwise neutrino conversion~\cite{2017arXiv170106580W}, and the effects of spin (helicity) coherence~\cite{2016arXiv161101862C}. Ref.~\cite{2016arXiv160705938F} discusses the trajectory-dependence of flavor transformation in these environments, and finds a variety of behaviors on different trajectories in antineutrino-dominated conditions, including MNR, synchronized MSW (Mikheyev-Smirnov-Wolfenstein) flavor transformation, and bipolar oscillations (for the inverted mass hierarchy). Here, we do a complementary study involving flavor evolution along a different trajectory (axial), and with different choices of parameters such as luminosities, spectra, and matter densities. Specifically, we focus on conditions where the total number luminosity (integrated number flux) of electron neutrinos is higher than that of electron antineutrinos, although we present results for antineutrino-dominated cases as well.

The geometry of a disk-like neutrino source, a \lq\lq neutrino disk\rq\rq, is mathematically more difficult to implement than a spherical source, i.e., a \lq\lq neutrino sphere\rq\rq. A disk-like geometry admits fewer symmetries than a spherical geometry, and thereby increases the degrees of freedom (by two, see Sec.~\ref{sec:Simulations}) that one needs to keep track of in order to fully self-consistently treat neutrino interactions in these environments. In order to keep the calculations tractable with the current technology available to us, we chose to run all simulations using a \lq\lq single-angle approximation\rq\rq\ (see Sec.~\ref{sec:Simulations}), and track the flavor evolution of neutrinos which stream out along the polar axis of the neutrino disk. Along this trajectory there is an azimuthal symmetry which we can exploit to make calculations simpler. Since we are tracking the flavor-histories of only the polar-axis directed neutrinos, any conclusions on $r$-process effects will apply only to the last of the three aforementioned $r$-process contributions in BNS mergers, i.e., the wind-like ejecta outside the equatorial disk plane.

Merger simulations that use different neutron star equations of state result in different initial conditions in terms of neutrino number luminosities and energy spectra. These differing initial conditions can then lead to qualitatively different flavor transformation phenomena, which can have implications for observables such as the amount and composition of ejecta, or the properties of the final remnant. An investigation of flavor transformation phenomena for these diverse initial conditions associated with different equations of state is therefore essential for accurately assessing the various possibilities across this landscape. Our results can be broadly separated into two classes, the matter-neutrino resonance (MNR) results for initial conditions where antineutrinos have higher number luminosities than neutrinos, and the bipolar spectral swap results for the neutrino-dominated number luminosities. We present results for both cases; however, we focus our discussion of flavor-transformation physics on the bipolar spectral swap results, since the MNR phenomenon has already been discussed extensively in this context~\cite{2016arXiv160705938F, 2014arXiv1403.5797M, 2016arXiv160704671Z, 2016PhRvD..93d5021M}. An example scenario where the neutrinos outnumber the antineutrinos can be found in Ref.~\cite{2016PhRvD..93d4019F}, in a merger simulation that uses the \lq\lq DD2\rq\rq\ equation of state for neutron star matter~\cite{0004-637X-748-1-70, PhysRevC.81.015803}. One aspect in which the DD2 equation of state differs from the other ones considered in Ref.~\cite{2016PhRvD..93d4019F}, is a higher degree of stiffness, and the associated high maximum cold neutron star mass limit, leading to a stable neutron star remnant even after spin-down.

In Sec.~\ref{sec:Method} we detail the method we used, both the mathematical model and the computational methods, to obtain our results. In Sec.~\ref{sec:results} we present our results. Sec.~\ref{sec:discussion} contains a discussion of the underlying physics of the flavor transformations, as well as of the likely implications of our results for the nucleosynthesis and other physics in these environments. We state our conclusions in Sec.~\ref{sec:conclusions}.

\section{Methodology}\label{sec:Method}

Neutrinos propagating in dense matter can change their flavors through both scattering-induced decoherence and through coherent, forward-scattering processes. This behavior is generally described by the quantum kinetic equations, essentially generalizations of the Boltzmann equation, but including quantum mechanical phases~\cite{Rudzsky1990, Sigl:1993fr, Raffelt:1993kx, McKellar:1994uq, Volpe:2013lr, 2013PhRvD..88j5009Z, Vlasenko:2014lr, 2015IJMPE..2441009V, 2015PhRvD..91l5020K}. However, in the regions where we see interesting flavor transformation effects, i.e., far above the BNS merger neutrino disk, treating only coherent forward scattering will likely be a good approximation. In fact, the conditions in these regions of interest along the polar axis resemble those of the supernova late-time neutrino-driven wind, which has been shown to be safe from neutrino halo effects~\cite{Cherry:2012lu}.

\subsection{Hamiltonian}\label{sec:Hamiltonian}
 
Here we treat only the coherent evolution of neutrino flavor. In this limit, neutrinos undergo forward scattering on background matter and on other neutrinos. The flavor state of a neutrino of energy $E_\nu$, at a location $r$ (for the axially directed trajectory, $r = z$) can be described by a two or three dimensional (depending on the number of flavors considered) state vector $|\Psi_{\nu}\rangle$, which evolves according to the Schr\"odinger-like equation~\cite{Wolfenstein78, Mikheyev85, Halprin:1986pn, Fuller87, Notzold:1988fv}:
\begin{equation}
i\hbar\frac{\partial}{\partial r}|\Psi_{\nu}(r,E_\nu)\rangle=H(r,E_\nu)|\Psi_{\nu}(r,E_\nu)\rangle.
\end{equation}

The Hamiltonian governing the evolution of these neutrinos is quite similar to those used in many previous simulations of collective neutrino flavor evolution~\cite{Duan06a, Duan06b, Duan06c, Duan07a, Duan07b, Duan07c, Duan08, Duan:2008qy, Duan:2008eb, Cherry:2010lr, Duan:2010fr, Duan:2011fk, Cherry:2011bg, Cherry:2012lu, 2013PhRvD..88j5009Z, 2014AIPC.1594..313B, 2015AIPC.1666g0001L, 2016AIPC.1743d0001B, 2016NuPhB.908..382A, 2016JCAP...01..028C, Barbieri:1991fj, 2016arXiv160906747V, Balantekin:2007kx, 2016PhRvD..94h3505J, Raffelt:2013qy, Sarikas:2012fk, Raffelt07, Hannestad06, PhysRevD.72.045003, PhysRevD.78.033014, PhysRevD.79.105003, Dasgupta09, Notzold:1988fv, Pastor:2002zl, Dasgupta:2008kx, Cherry:2013lr, Cherry:2012lr, 2015PhRvD..92l5030D, 2015PhLB..751...43A, 2015PhLB..747..139D, 2015IJMPE..2441008D, 2015PhRvD..92f5019A, 2014JCAP...10..084D, Qian93, 1992AAS...181.8907Q, Qian95, 1995PhRvD..52..656Q, 2011PhRvD..84e3013B}. In particular, we used a version of the Hamiltonian from the \lq\lq neutrino bulb\rq\rq\ model used in supernova neutrino flavor evolution studies~\cite{Duan:2010fr, Duan06a}, modified to suit a BNS merger disk geometry. For the two-flavor case, easily generalizable to three flavors, the Hamiltonian is~\cite{Duan06a,Duan06c,Duan:2010fr,1989neas.book.....B,2014AIPC.1594..313B}:
\begin{widetext}
\begin{equation}
\begin{aligned}
H(r,E_\nu) = \frac{\delta m^2}{4E_\nu}U\begin{bmatrix}-1 & 0 \\ 0 & 1\end{bmatrix}U^\dagger 
			+ \sqrt{2}G_Fn_e(r)\begin{bmatrix} 1 & 0 \\ 0 & 0\end{bmatrix} 
	&+ \sqrt{2}G_F\sum_\alpha\int_\nu dn_{\nu,\alpha}(\bm{p}')\left|\Psi_{\nu,\alpha}(\bm{p}')\right>\left<\Psi_{\nu,\alpha}(\bm{p}')\right|(1-\hat{p}\cdot\hat{p}') \\ 
	&- \sqrt{2}G_F\sum_\alpha\int_{\bar\nu} dn_{\bar\nu,\alpha}(\bm{p}')\left|\Psi_{\bar\nu,\alpha}(\bm{p}')\right>\left<\Psi_{\bar\nu,\alpha}(\bm{p}')\right|(1-\hat{p}\cdot\hat{p}'),
\label{eqn:Hamiltonian}
\end{aligned}
\end{equation}
\end{widetext}
where $U$ is the $2 \times 2$ version of the Pontecorvo-Maki-Nakagawa-Sakata (PMNS) mixing matrix:
\begin{equation}
U=\begin{bmatrix}\cos\theta_v & \sin\theta_v \\ -\sin\theta_v & \cos\theta_v\end{bmatrix},
\label{eqn:PMNS}
\end{equation}
with $\theta_v = 8.7^\circ$ the mixing angle in vacuum (we have used the $\theta_{13} \simeq 8.7^\circ$~\cite{Olive:2016xmw} mixing angle in our $2 \times 2$ simulations). Here $\delta m^2 = 2.4\times 10^{-3}\, \text{eV}^2$ is the mass-squared splitting (we have used the atmospheric splitting), $G_F$ is the Fermi weak coupling constant, $n_e$ is the net electron number density ($n_e = n_{e^-} - n_{e^+}$), and $\bm{p} ,\, \bm{p}'$ are the momenta of the test and background neutrinos, respectively ($E_\nu = \sqrt{\bm{p}^2 + m_\nu^2}$). We integrate over all of the background neutrinos encountered by our test neutrino, so that $dn_{\nu,\alpha}(\bm{p}')$ is the local number density of neutrinos in state $\left|\Psi_{\nu,\alpha}(\bm{p}')\right>$. Here, the index \lq\lq $\alpha$\rq\rq\ refers to the initial flavor in which the neutrino was emitted \textit{at the neutrino disk}.

The three terms of the Hamiltonian in Eq. (\ref{eqn:Hamiltonian}) are written in the order of the vacuum Hamiltonian ($H_\text{vac}$), the matter Hamiltonian ($H_m$), and the neutrino-neutrino \lq\lq self coupling\rq\rq\ Hamiltonian ($H_{\nu\nu}$). The vacuum term in Eq. (\ref{eqn:Hamiltonian}), $H_\text{vac}$ arises merely from the fact that neutrinos have mass, and that the mass eigenstates are not coincident with the neutrino flavor eigenstates. The matter term, $H_m$, arises from the neutrino forward scattering via the charged current interactions on the background matter (the potential from neutral current interactions contributes equally to all flavors of neutrinos and therefore does not need to be considered here). This matter Hamiltonian depends on the electron number density $n_e$, which can be written in terms of the baryon number density $n_b$ and the electron fraction $Y_e$:
\begin{equation}
n_e=Y_e n_b \, .
\end{equation}
We chose the baryon density profile to have an inverse cubic relation to the radius (distance from the disk): 
\begin{equation}
n_b = n_{b,0} \left(\frac{r_0}{r}\right)^3
\end{equation}
where $n_{b,0}$ is the initial baryon density at the initial radius $r_0$. This relation will hold true if the material is in hydrostatic equilibrium and the entropy is mostly carried by relativistic particles (see Ref.~\cite{Duan06a, 1996ApJ...471..331Q}). The last term in the Hamiltonian, $H_{\nu\nu}$, in Eq. (\ref{eqn:Hamiltonian}) arises from the test neutrino forward-scattering on other background neutrinos. This is the term which depends on the geometry which we chose, and requires careful consideration.  

First, by exploiting the azimuthal symmetry of our chosen trajectory, we can rewrite the expression $(1-\hat{p}\cdot\hat{p}')$ in a convenient form:
\begin{equation}
1-\hat{p}\cdot\hat{p}'  = 1-\cos\theta',
\label{eqn:phat}
\end{equation}
where the test neutrino trajectory is taken to be along the $z$-direction, and therefore, the intersection angle between the test and background neutrino trajectories is simply the polar angle $\theta'$ of the latter.

Second, the neutrino states can be enumerated in terms of energies and the pencil of solid angle (in the direction of $\bm{p}'$), subtended at our test neutrino's location, in which the neutrinos are streaming. Or in more concrete mathematical terms:
\begin{equation}
dn_{\nu,\alpha}=\frac{N_{\nu,\alpha}}{2\pi^2}\left(\frac{d\Omega_\nu}{4\pi}\right)f_{\nu,\alpha}(E_\nu)dE_\nu \, .
\label{eqn:dn}
\end{equation} 

Here, $N_{\nu,\alpha}$ is a factor that normalizes the number density to the energy luminosity $L_{\nu,\alpha}$ in the respective neutrino flavor. $f_{\nu,\alpha}(E_\nu)$ is the (non-normalized) energy distribution of neutrinos initially emitted in state $\alpha$ and $d\Omega_\nu$ is a differential solid angle. We assume that neutrinos are emitted from the surface of the neutrino disk with a Fermi-Dirac black body-shaped distribution of energies so that:
\begin{equation} 
f_{\nu,\alpha}(E_\nu)=\frac{E_\nu^2}{e^{E_\nu/T_{\nu,\alpha}-\eta_{\nu,\alpha}}+1} \, ,
\end{equation}
where $T_{\nu,\alpha}$ and $\eta_{\nu,\alpha}$ are the temperature and degeneracy parameter, respectively, of the initial $\nu_\alpha$ distribution. To normalize the differential number density $dn_{\nu,\alpha}$ with respect to the luminosity $L_{\nu,\alpha}$, we first calculate the neutrino energy flux $F_{\nu,\alpha}$ at the disk surface:
\begin{equation}
\begin{aligned}
F_{\nu,\alpha} &= \int_{\bm{p}}  dn_{\nu,\alpha}  E_\nu \cos\theta \\
			&= \frac{N_{\nu,\alpha}}{2\pi^2}\int\displaylimits_0^{2\pi}\int\displaylimits_0^{1}\cos\theta\frac{d\cos\theta \, d\phi}{4\pi} \int\displaylimits_0^\infty E_\nu f_{\nu,\alpha}(E_\nu)dE_\nu,
\end{aligned}
\end{equation}
where the neutrino speed is taken to be the speed of light $c = 1$, and the angle integration is performed over half the sky, i.e., over all neutrino unit momenta on one side of the disk. Now, we can introduce the normalized energy distribution function $\tilde{f}_{\nu,\alpha}(E_\nu)$ defined as:
\begin{equation}
\tilde{f}_{\nu,\alpha}(E_\nu)=\frac{1}{T_{\nu,\alpha}^3 F_2(\eta_{\nu,\alpha})}\frac{E_\nu^2}{e^{E_\nu/T_{\nu,\alpha}-\eta_{\nu,\alpha}}+1},
\end{equation}
so that $\int_0^\infty \tilde{f}_{\nu,\alpha}(E_\nu)dE_\nu = 1$, where $F_2(\eta_\nu)$ is the complete Fermi-Dirac integral of order 2. With this, we can then evaluate the above integral to obtain
\begin{equation}
F_{\nu,\alpha}=\frac{N_{\nu,\alpha}}{8\pi^2} \langle E_{\nu,\alpha} \rangle \, T_{\nu,\alpha}^3 \, F_2(\eta_{\nu,\alpha}),
\label{eqn:flux2}
\end{equation}
where $\langle E_{\nu,\alpha} \rangle$ is the average neutrino energy over the distribution $\tilde{f}_{\nu,\alpha}$. We can now fix $N_{\nu,\alpha}$ by relating the flux to the luminosity using $F_{\nu,\alpha}={{L_{\nu,\alpha}}/{(2\pi R_\nu^2)}}$, giving us 
\begin{equation}\label{eq:dn}
dn_{\nu,\alpha}=\frac{L_{\nu,\alpha}}{2\pi^2 R^2_\nu\langle E_{\nu,\alpha}\rangle}\tilde{f}_{\nu,\alpha}(E_{\nu,\alpha})d\Omega_\nu dE_\nu \, .
\end{equation}

Note that this differs by a factor of two from the normalization for a spherical emission geometry. With this normalization, we can write the neutrino-neutrino Hamiltonian as an explicit integral:
\begin{widetext}
\begin{equation}
H_{\nu\nu}=\frac{\sqrt{2}G_F}{\pi R^2_\nu}\sum_\alpha\int\displaylimits_0^\infty\int\displaylimits_0^{\theta_m}\left[\frac{L_{\nu,\alpha}}{\langle E_{\nu,\alpha}\rangle}\tilde{f}_{\nu,\alpha}(E_\nu)\left|\Psi_{\nu,\alpha}\right>\left<\Psi_{\nu,\alpha}\right| - \frac{L_{\bar\nu,\alpha}}{\langle E_{\bar\nu,\alpha}\rangle}\tilde{f}_{\bar\nu,\alpha}(E_\nu)\left|\Psi_{\bar\nu,\alpha}\right>\left<\Psi_{\bar\nu,\alpha}\right| \right] (1-\cos\theta')\sin\theta'  d\theta' dE_\nu.
\label{eqn:Hnunu}
\end{equation} 
\end{widetext}
Here $\theta_m$ is the maximum half-angle the neutrino disk subtends at the test neutrino's location which with simple trigonometry we know to be: $\tan\theta_m=R_\nu/r$. We have already performed the $\phi'$ integration in Eq.~(\ref{eqn:Hnunu}) as that integral is trivially equal to $2\pi$ because of azimuthal symmetry. This is the final version of the Hamiltonian which we use in our calculations. We can see that the main difference between this Hamiltonian and the one used in \cite{Duan06a} will be in how $\theta_m$ differs between the spherical geometry case and the disk geometry case. As $\theta_m$ is different between the two cases, the geometric dilution of neutrinos as we move farther from the source will be different.

\subsection{Simulations} \label{sec:Simulations}

Here we chose to model the neutron star merger neutrino source as a flat circular disk, with neutrinos streaming from the two faces. We assumed that all neutrinos of all flavors are emitted from the same surface, i.e., that there are not multiple neutrino disks for different neutrino flavors. Neither of these assumptions are quite true for an actual neutron star merger; different neutrino flavors and types have different decoupling surface disks, and these have relative spacings of order tens of km, at most. Differences in neutrino decoupling surfaces for different neutrino types has been shown to be important~\cite{PhysRevLett.116.081101, 2017arXiv170106580W}, and that could be the case here as well, especially close to the neutrino disk. However, as we shall see, most of the collective flavor oscillations we find occur at distances ($\sim$ a few hundred km) which are large compared to the neutrino disk separations. The effects of having separate disks are therefore unlikely to be significant at these distances.

We chose a disk radius of $R_\nu = 60 \, \text{km}$ (see, e.g., Fig. 16 in Ref.~\cite{2016PhRvD..93d4019F}), and assumed that neutrinos are emitted isotropically from each point on the surface. Moreover, as mentioned earlier, we chose to follow the flavor evolution of neutrinos emitted from the center of the disk along the polar axis, perpendicular to the equatorial plane. 

Simulations were performed using the neutrino BULB code, developed by the authors of Refs.~\cite{Duan06a, Duan06b, Duan06c, Duan07a, Duan07b, Duan07c, Duan08, Duan:2008qy, Duan:2008eb, Cherry:2010lr, Duan:2010fr}. The BULB code was modified to use the new geometry as discussed in Sec.~\ref{sec:Hamiltonian}. No major modifications to the underlying architecture of the code were necessary. We note that Eq. (\ref{eqn:Hnunu}) was written in such a way as to leave the angle dependence of $|\Psi_{\nu,\alpha}\rangle$ ambiguous. In a fully self-consistent study of a merger geometry, this state vector would of course depend on the trajectory of the background neutrino we are integrating over. Indeed, even in a spherical \lq\lq neutrino bulb\rq\rq\ geometry, this state vector is emission angle (angle with respect to the normal of the neutrino sphere at which a particular neutrino is emitted) dependent. So called multi-angle simulations of the neutrino bulb model account for this fact. However, with the disk geometry, there are two additional degrees of freedom, beyond multi-angle bulb simulations, that we must account for if we want to fully self consistently treat the neutrino trajectories. As the disk is not spherically symmetric, we must account for the emission location on the disk (one degree of freedom due to cylindrical symmetry). In addition, multi-angle bulb simulations only require one polar emission angle whereas a disk geometry would require two emission angles (polar and azimuthal) to track all of the neutrinos. Complicating matters further, as off-angle trajectories (from the central polar axis) do not exhibit azimuthal symmetry, the relation in Eq. (\ref{eqn:phat}) no longer holds. Also, for off-axis trajectories, the integral over the solid angle would depend on the polar angle $\theta$ in addition to $r$, and the separation into $\theta'$ and $\phi'$ integrals is nontrivial. As such, the underlying architecture of the BULB code would have to be modified to accommodate these extra degrees of freedom.

To avoid these complexities, simulations were run for this paper in the so-called \lq\lq single-angle mode\rq\rq. We assume, for simplicity, that all neutrinos on all other trajectories that encounter our test neutrino evolve in flavor exactly the same way as our test neutrino. It is known from simulations of supernova neutrino flavor evolution that multi-angle simulations incorporate the correct phase-averaging over different trajectories, implying that they can do a better job at predicting the locations of the onset of collective flavor transformations~\cite{Duan:2011lr}. Nevertheless, single-angle simulations are known to capture many of the \textit{qualitative} features that are present in multi-angle simulations, especially at locations sufficiently far from the source. We also note that previous studies of flavor evolution in BNS merger environments have also employed this approximation.

Since we are performing single-angle simulations, the state vectors $|\Psi_{\nu,\alpha}\rangle$ are not emission location or angle dependent. They are, however, still energy dependent. As such, we can perform the angle integration in Eq. (\ref{eqn:Hnunu}) to obtain:
\begin{widetext}
\begin{equation}
H_{\nu\nu}= \frac{\sqrt{2}G_F}{\pi R^2_\nu}\cdot\frac{1}{2}\left[1-\frac{r}{\sqrt{r^2+R_\nu^2}}\right]^2\sum_\alpha\int\displaylimits_0^\infty \left[\frac{L_{\nu,\alpha}}{\langle E_{\nu,\alpha}\rangle}\tilde{f}_{\nu,\alpha}(E_\nu)\left|\Psi_{\nu,\alpha}\right>\left<\Psi_{\nu,\alpha}\right| - \frac{L_{\bar\nu,\alpha}}{\langle E_{\bar\nu,\alpha}\rangle}\tilde{f}_{\bar\nu,\alpha}(E_\nu)\left|\Psi_{\bar\nu,\alpha}\right>\left<\Psi_{\bar\nu,\alpha}\right| \right] dE_\nu.
\end{equation}  
\end{widetext}
    
\section{Results}\label{sec:results}

\subsection{Initial Conditions}

As the merger environment itself is extremely complex, the neutrino emission's initial conditions are, not surprisingly, equally complex. The major regions of neutrino emission differ for the different neutrino flavors, and among the neutrino and the antineutrino sector. Most importantly with regards to the flavor transformations, neutrinos are emitted primarily from the polar regions of the merger while antineutrinos are mostly emitted from the hot shocked regions of the disk~\cite{2016PhRvD..93d4019F}. This means that different simulations giving differing temperatures for the polar regions versus the shocked regions of the disk, would give similarly different results in neutrino versus antineutrino emission. Most nuclear matter equations of state in use in BNS merger simulations result in a higher luminosity and number flux of antineutrinos over neutrinos being emitted from the neutrino disk~\cite{2009ApJ...690.1681D, 2014MNRAS.443.3134P, 1997A&A...319..122R, 2016PhRvD..93d4019F}. However, a particular simulation from Ref.~\cite{2016PhRvD..93d4019F}, one that used the \lq\lq DD2\rq\rq\ equation of state for neutron matter, did produce a total number luminosity abundance of neutrinos over antineutrinos (although, due to the high average energy of the antineutrinos, the energy luminosity was still dominated by antineutrinos). 

We do not include all of the intricacies of neutrino emission from the neutrino disk. Instead, we chose different sets of initial neutrino luminosities and energy spectra in order to try to capture the qualitative differences in flavor evolution which arise from differences in initial conditions. As most simulations of neutrino emission have antineutrino dominance, studies of flavor evolution in merger environments up to now have focused on the \lq\lq matter-neutrino resonance (MNR)\rq\rq\ effect~\cite{2016arXiv160705938F, 2014arXiv1403.5797M, 2016arXiv160704671Z, 2016PhRvD..93d5021M}. This effect requires a cancellation in the total Hamiltonian between the matter term and the neutrino-neutrino term. Such a cancellation can only arise if the neutrino-neutrino term is negative, i.e., if the neutrinos are dominated by antineutrinos. To corroborate this, we ran a simulation with the same neutrino luminosities and spectra as found in Ref.~\cite{2016arXiv160705938F}, with an antineutrino abundance over neutrinos, and found that the MNR effect was indeed the dominant feature of collective neutrino oscillations. However, if neutrinos dominate over antineutrinos in number flux, the matter-neutrino resonance cannot easily occur\footnote{To get a MNR in such a scenario, one would need other mechanisms to first convert some of the electron neutrino lepton number excess, either into other flavors (e.g., via background-assisted MSW effect), or into anti-neutrinos (e.g., $\nu_e \rightarrow \bar\nu_e$ via spin-coherence effects~\cite{2017PhRvD..95f3004T, 2014arXiv1406.6724V, 2016arXiv161101862C}).}. In Secs.~\ref{sec:2fbipolarresult}--\ref{lowlumlowdens}, we highlight a different possible outcome of collective flavor oscillations in a merger environment, namely, the occurrence of a bipolar spectral swaps for neutrino-dominated number luminosities.


The simulations, the results of which are described below, all utilize the normal neutrino mass hierarchy. We take $n_{b,0} = 10^8 \, \text{g/cm}^3$ (at $r_0 = 20$ km), which is the same as in Ref.~\cite{2016PhRvD..93d4019F}, in all our simulations except for the low luminosity/low density simulation of Sec.~\ref{lowlumlowdens} which uses $n_{b,0} = 2.5 \times 10^6 \, \text{g/cm}^3$. In addition, we chose a constant electron fraction $Y_e = 0.4$ in our calculations. All simulations reported in this work were performed using the single angle approximation for calculating neutrino flavor evolution. Examining previous works, we can see that this approximation has been found to capture with fair fidelity the qualitative behavior of this evolution in most supernova environments, but it is known to fail quantitatively in some cases (see section \ref{sec:Simulations} for details). Consequently, we caution that our single angle simulations may give results which differ from a full multi-angle treatment.

\subsection{MNR Results}\label{sec:MNR}

As outlined above, we will have the requisite conditions for MNR when $((L_{\nu,e}/\langle E_{\nu,e}\rangle)/(L_{\bar{\nu},e}/\langle E_{\bar{\nu},e}\rangle))<1$. In MNR, the neutrino-neutrino part of the Hamiltonian in Eq. (\ref{eqn:Hamiltonian}) interacts with the matter and vacuum parts of the Hamiltonian to create an MSW-like resonance~\cite{Qian95, 2016arXiv160705938F, 2014arXiv1403.5797M, 2012PhRvD..86h5015M, 2016PhLB..752...89W, 2016PhRvD..93d5021M, 2016PhRvD..93j5044V, 2016arXiv160504903S, 2016arXiv160704671Z}. A resonance between the two flavors of neutrinos occurs when the diagonal elements of the total Hamiltonian equal each other, i.e. when $H_{11}=H_{22}$. As is standard in neutrino flavor evolution analyses, we use a traceless Hamiltonian in our simulations by removing the total trace from Eq.~(\ref{eqn:Hamiltonian}). For a traceless $2\times 2$ Hamiltonian, the resonance condition is then simply $H_{11}=-H_{22}=0$. A MNR therefore occurs when $(H_{\nu\nu})_{11}$ nearly cancels $(H_m)_{11}$. Usually, MNR is augmented by nonlinear feedback in the neutrino flavor evolution which helps sustain this cancellation over a longer distance. Generally speaking, if the neutrinos move through this MNR adiabatically, then large scale flavor transformations can occur from the electron flavor state to the \lq\lq $x$\rq\rq- flavor state and vice versa. Here, the $x$-flavor refers to the other flavor besides $\nu_e$ in $2\times2$ calculations and is taken to be a particular linear combination of the nearly maximally mixed $\nu_\mu$ and $\nu_\tau$ flavor states~\cite{Balantekin:2000hl,Caldwell:2000db}.

In the normal mass hierarchy, the vacuum Hamiltonian matrix element $(H_\text{vac})_{11}$ is an energy dependent negative quantity. However, since $H_m$ and $H_{\nu\nu}$ are not energy dependent, the MNR cannot simultaneously satisfy $H_{11}=0$ for neutrinos of all energies. The diagonal Hamiltonian can therefore vanish only for one specific energy, and it can be close to zero only for neutrinos with energies close to that energy. In other words, not all neutrinos of all energies may necessarily be affected by the MNR. This general observation is borne out in our simulations.    

Table \ref{table:MNRparams} shows the parameters we used in order to simulate neutrino flavor evolution using an antineutrino dominated neutrino number luminosity. These parameters for the neutrino luminosities and average energies are quite similar to those used in~\cite{2016arXiv160705938F}. Notice that here  $((L_{\nu,e}/\langle E_{\nu,e}\rangle)/(L_{\bar{\nu},e}/\langle E_{\bar{\nu},e}\rangle))\approx 0.7$ which means there is a preponderance of antineutrinos over neutrinos and therefore the possibility for a MNR. 

\begin{table}[ht]
\begin{ruledtabular}
\begin{tabular}{lcdr}
\text{Parameter} & 
\text{Value} \\
\colrule
	$L_{\nu,e}$ & $1.5\times 10^{52}\, \text{erg/s}$\\
	$L_{\bar\nu,e}$ & $3.0\times 10^{52}\, \text{erg/s}$ \\
   	$L_{\nu,\bar\nu,x}$ & $1.6\times 10^{52}\,\text{erg/s}$\\
	$\langle E_{\nu,e}\rangle$ & $10.6\, \text{MeV}$\\
	$\langle E_{\bar\nu,e}\rangle$ & $15.3\, \text{MeV}$\\
	$\langle E_{\nu,\bar\nu,x}\rangle$ & $17.3\, \text{MeV}$\\
	$\delta m^2_\text{atm}$ & $2.4\times 10^{-3}\, \text{eV}^2$\\
	$\theta_\text{V}$ & $8.7^\circ$\\
\end{tabular}
\caption{Parameters used our two flavor simulation that produced MNR. The luminosities and average energies used here are taken from \cite{2016arXiv160705938F}.}
\label{table:MNRparams}
\end{ruledtabular}
\end{table}

\begin{figure*}[ht]
\centering
\subfloat{\includegraphics[width=.5\textwidth]{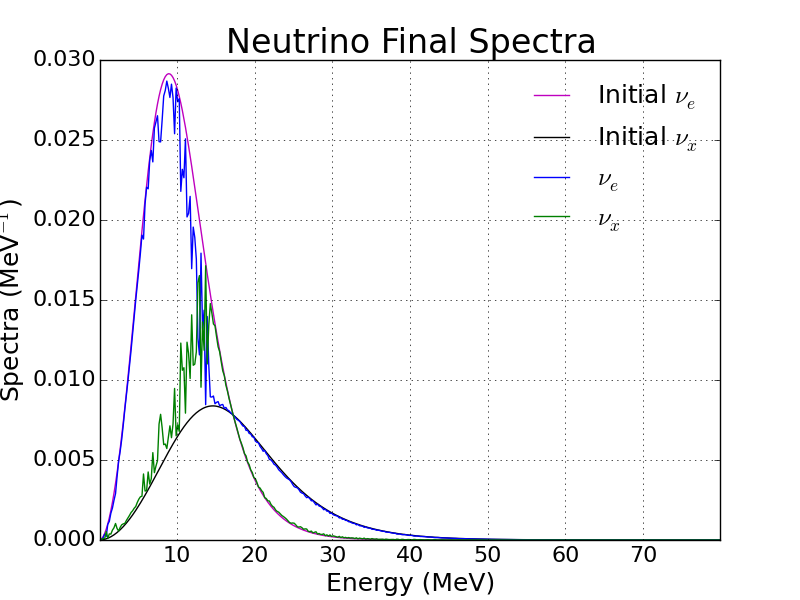}}
\subfloat{\includegraphics[width=.5\textwidth]{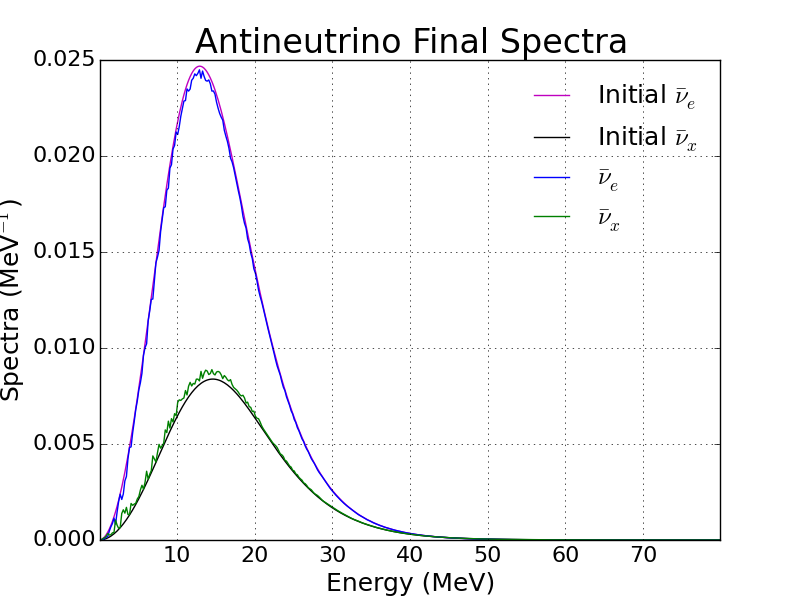}}
\caption{These plots show the initial (magenta and black) and final (blue and green) energy spectra for neutrinos (left) and antineutrinos (right), for the simulation with parameters as described in Table~\ref{table:MNRparams}. The final spectra were plotted at a distance $r = 5000$ km along the polar axis. As is evident, the MNR affected primarily the high-energy neutrinos. Neutrinos with energies below $E_\nu \lesssim 16 \, \text{MeV}$ and antineutrinos were not significantly affected.}
\label{fig:MNRspectra}
\end{figure*}

Figure \ref{fig:MNRspectra} shows the energy spectra for neutrinos and antineutrinos at a final simulation radius of $5000 \, \text{km}$, along with the initial spectra at the point of emission. These neutrinos moved through an MNR as shown in Fig.~\ref{fig:MNRHamiltonian}. We can see here that only the high energy neutrinos converted from one flavor to the other, while low energy neutrinos did not change flavors. This stems from the fact that the MNR set $H_{11}\approx 0$ only for these high energy neutrinos. 

For a neutrino of energy $E_\nu$ emitted initially in the $\alpha$ flavor state, the probability of being in the $\beta$ flavor state at a distance $r$ is $P_{\alpha\beta}(r,E_\nu) = \vert \langle \nu_\beta \vert \Psi_{\nu,\alpha} (r,E_\nu) \rangle \vert^2$. This can then be integrated over neutrino energies, weighted by the normalized distribution functions $\tilde{f}_{\nu,\alpha}(E_\nu)$, to obtain the energy-averaged survival ($\alpha = \beta$) or conversion ($\alpha \neq \beta$) probability as a function of distance:
\begin{equation}
P^\text{avg}_{\alpha\beta}(r) = \int \tilde{f}_{\nu,\alpha}(E_\nu) P_{\alpha\beta}(r,E_\nu) dE_\nu,
\end{equation}

\begin{figure*}[ht]
\centering
\subfloat{\includegraphics[width=.5\textwidth]{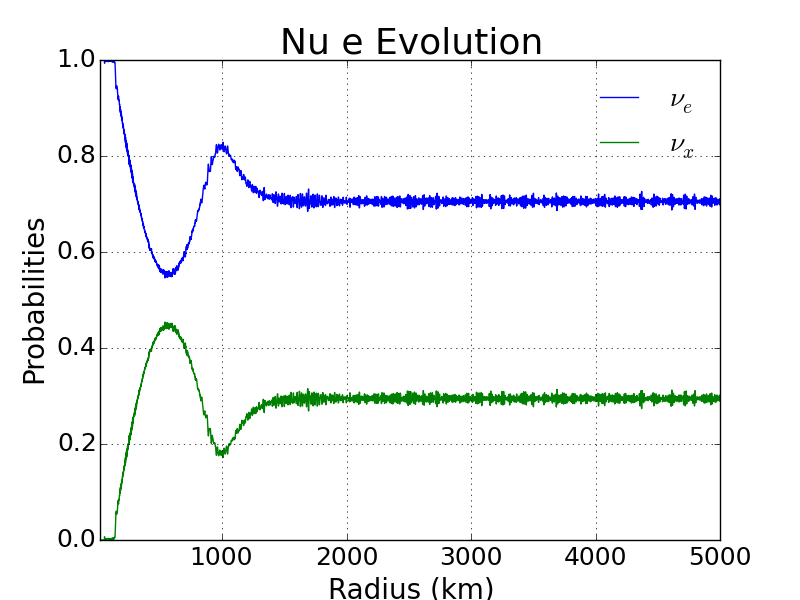}}
\subfloat{\includegraphics[width=.5\textwidth]{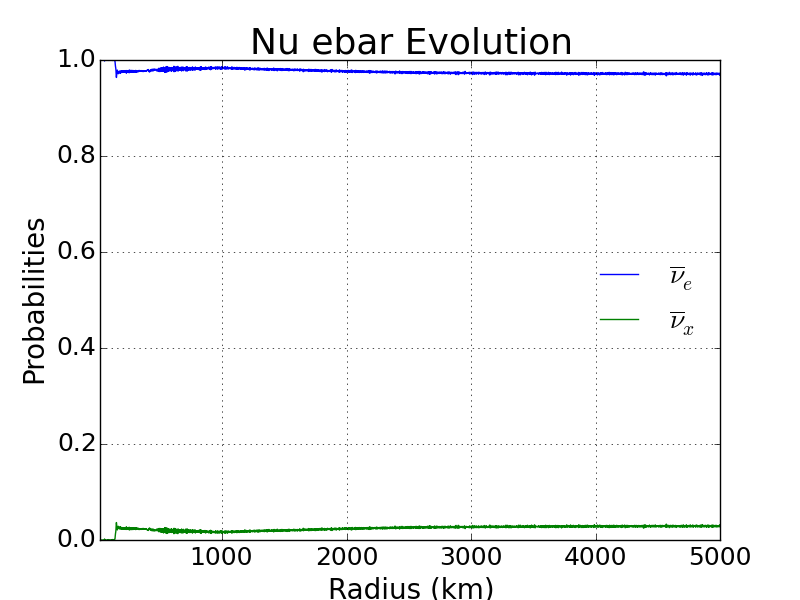}}
\caption{These figures show the energy-averaged flavor evolution of a neutrino (left) and antineutrino (right), initially in the electron flavor state, for the simulation with parameters as described in Table~\ref{table:MNRparams}. It is evident that the MNR begins early on at a distance of about $\approx 200 \, \text{km}$ and stabilizes at about $\approx 1200 \, \text{km}$. In each of these energy-averaged flavor evolution plots, the lines corresponding to different flavors (e.g., the blue and green lines in the left panel) sum to unity at each radius.} 
\label{fig:NueEvolutionMNR}
\end{figure*}

\begin{figure*}[ht]
\centering
\subfloat{\includegraphics[width=.5\textwidth]{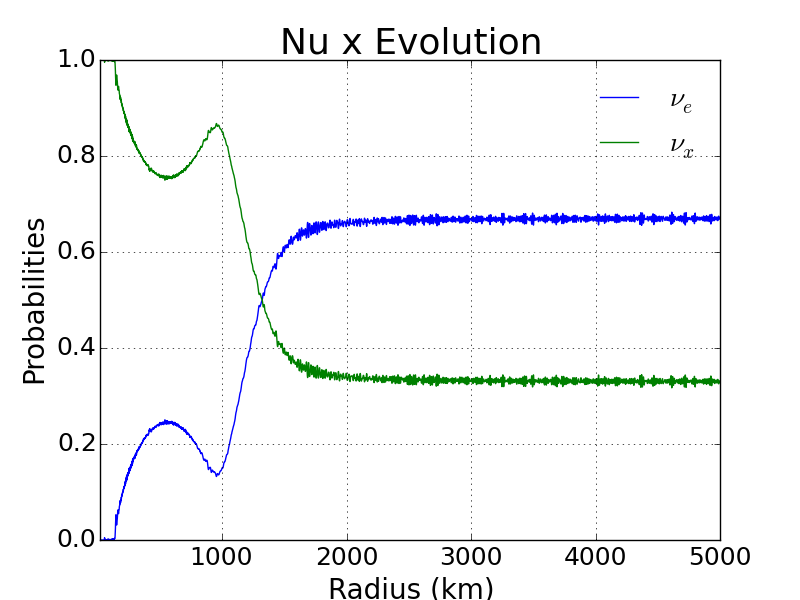}}
\subfloat{\includegraphics[width=.5\textwidth]{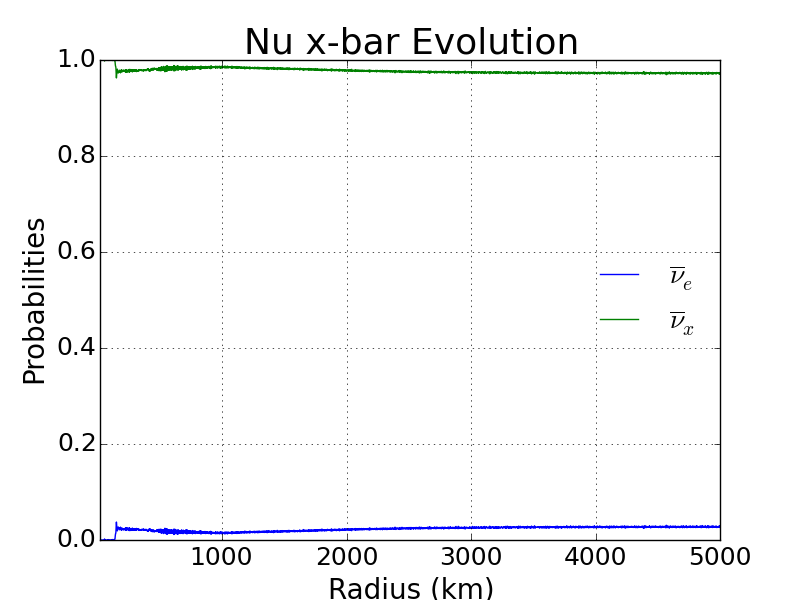}}
\caption{Same as Fig.~\ref{fig:NueEvolutionMNR}, but for a neutrino (left) and antineutrino (right) initially in the $x$ flavor state. Mirroring the results of the initially electron flavor neutrinos, most of the flavor transformation occurs for the neutrinos, while the antineutrinos remain largely unaffected by the MNR.} 
\label{fig:NuxEvolutionMNR}
\end{figure*}

\begin{figure}[ht]
\centering
\includegraphics[width=.5\textwidth]{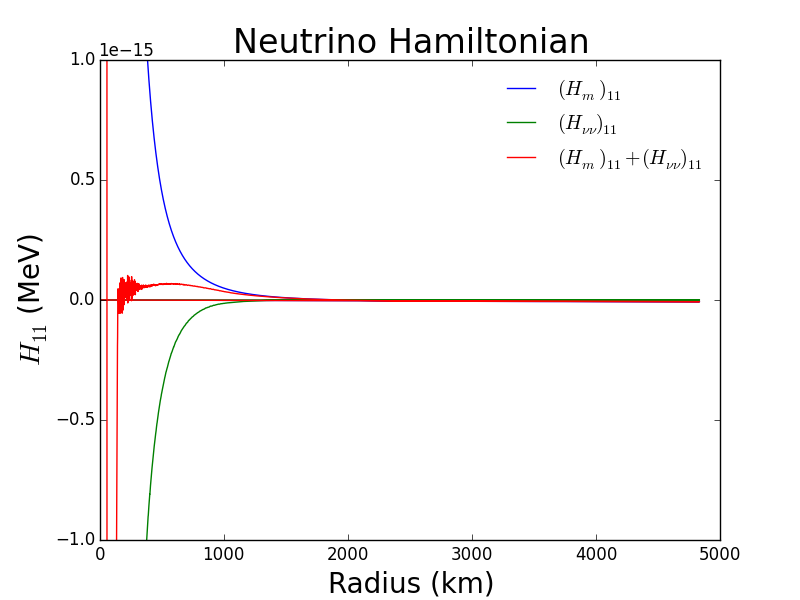}
\caption{Shown here is the 1-1 component of the matter, neutrino-neutrino, and the sum of the matter and neutrino-neutrino Hamiltonian experienced by our test neutrinos, for the simulation with parameters as described in Table~\ref{table:MNRparams}. As we can see, the MNR develops early on at a radius of $\approx 200 \, \text{km}$. The nonlinear feedback of neutrino flavor transformations keep the total Hamiltonian near 0 for several hundred kilometers, thus giving rise to the MNR. In order to calculate the total Hamiltonian, an energy dependent $(H_\text{vac})_{11}$ would have to be added. This vacuum term would manifest as an energy dependent vertical offset (in the negative direction). Thus, not all neutrinos of all energies will go through the MNR, explaining the energy dependence of the MNR effect seen in Fig.~\ref{fig:MNRspectra}.} 
\label{fig:MNRHamiltonian}
\end{figure}

Figure \ref{fig:NueEvolutionMNR} shows the energy-averaged flavor evolution probabilities for a neutrino and antineutrino which begin initially in the electron flavor state. As is evident, electron neutrinos began to convert into $x$-flavor neutrinos beginning quite close to the neutrino disk, at a distance of $\approx 200 \, \text{km}$. The collective flavor transformation ended by about $\approx 1200 \, \text{km}$. Figure \ref{fig:NuxEvolutionMNR} shows the energy-averaged flavor evolution of neutrinos and antineutrinos which begin in the $x$-flavor state. As can be concluded from Figs.~\ref{fig:NueEvolutionMNR} and \ref{fig:NuxEvolutionMNR} the antineutrinos did not significantly change flavors and only the neutrinos were affected by the MNR.

Figure \ref{fig:MNRHamiltonian} shows the 1-1 component of the matter and neutrino-neutrino Hamiltonians and the sum of the two for the simulation presented here. We can see that the nonlinear feedback from the MNR forced $(H_m)_{11}+(H_{\nu\nu})_{11} \approx 0$ over roughly a thousand kilometers ($r\approx 200\text{--} 1200 \, \text{km}$). The radius at which the MNR is reached essentially corresponds to the radius at which the neutrinos begin to transform their flavor, as seen from Figs.~\ref{fig:NueEvolutionMNR} and \ref{fig:NuxEvolutionMNR}, establishing that the MNR was indeed the mechanism driving flavor transformation in this simulation.

\begin{figure*}[ht]
\centering
\subfloat{\includegraphics[width=.5\textwidth]{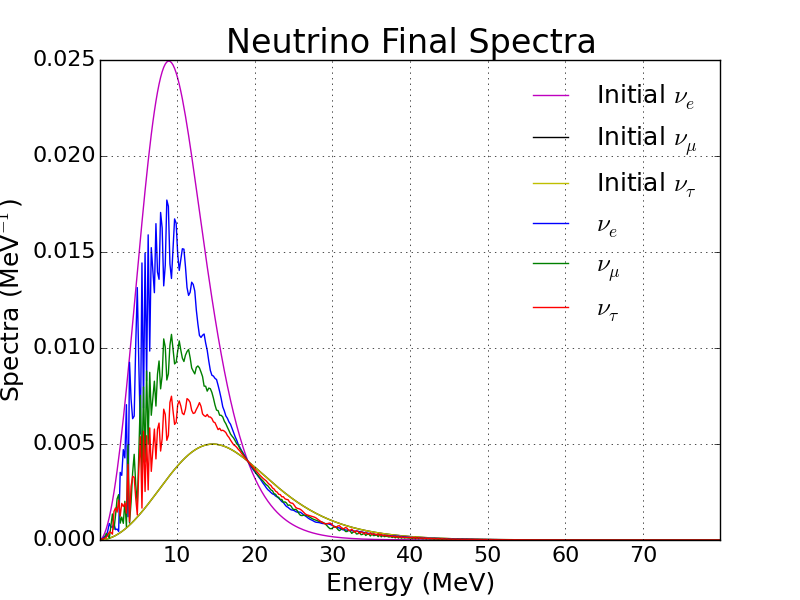}}
\subfloat{\includegraphics[width=.5\textwidth]{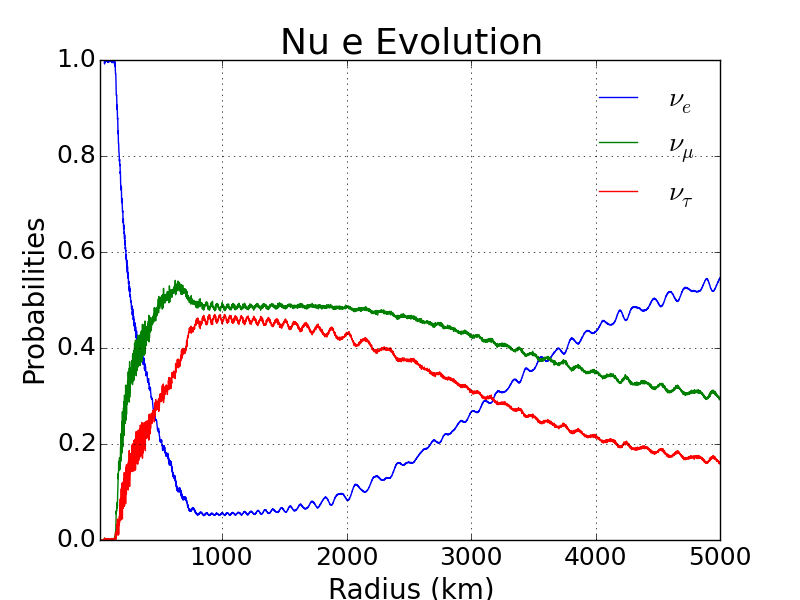}}
\caption{These figures show some of the results from a three-flavor MNR calculation with the same parameters as in Table~\ref{table:MNRparams}. (Left) Initial and final ($r = 5000$ km) neutrino spectra for a 3-flavor MNR calculation. (Right) Evolution of a neutrino initially in the electron flavor state.} 
\label{fig:3fMNR}
\end{figure*}

For comparison, we also performed a three-flavor calculation for this anti-neutrino dominated case with the same initial conditions as those in Table~\ref{table:MNRparams}, with the \lq\lq $x$\rq\rq-flavor luminosity equally split between the $\mu$ and $\tau$ flavors. Figure~\ref{fig:3fMNR} shows plots of initial and final neutrino spectra (left), as well as the energy-averaged flavor evolution of neutrinos starting out in the electron flavor state. As can be seen from the figure, the excursions in flavor space for the three-flavor calculation are bigger compared to the two-flavor case, and the collective effects do not die down completely, even by $r = 5000$ km, an effect that can be attributed to the influence of oscillations driven by the solar mass-squared splitting.

\subsection{Two Flavor Bipolar Swap Results}\label{sec:2fbipolarresult}

The results of the simulations where the initial number luminosities are dominated by neutrinos rather than antineutrinos can be quite different. Our choice of parameters that correspond to such a neutrino dominance over antineutrinos is motivated by Foucart \textit{et al.}'s DD2 equation of state simulation~\cite{2016PhRvD..93d4019F}.

We performed two- and three-flavor simulations in which we observed the bipolar spectral swap phenomenon. Table \ref{table:params} outlines the parameters used in the two flavor simulation discussed in this section. The parameters chosen here represent an example set of neutrino luminosities and spectra that one might expect in these environments, based on physical insight. For instance, in a neutron-rich environment one would expect a pronounced hierarchy between the average energies of $\nu_e$ and $\bar\nu_e$, as well as those of $\nu_e$ and $\nu_x$. This is because only the electron neutrinos would experience significant charged-current interactions, and would therefore be expected to decouple further out where the temperatures are cooler. In addition, the parameter set that we have chosen here also has a prominent hierarchy between $\bar\nu_e$ and $\nu_x$ average energies.

The rationale behind this choice was to explore a scenario wherein flavor transformations could significantly affect the nucleosynthesis prospects. This is discussed in further detail in Sec.~\ref{sec:ye}. Another justification is that there do exist simulations where such a hierarchy between $\bar\nu_e$ and $\nu_x$ average energies has been exhibited. For instance, neutrino emission from the \lq\lq SFHo\rq\rq\ equation of state simulation from Ref.~\cite{2016PhRvD..93d4019F} has average energies $\langle E_{\bar\nu,e}\rangle = 19.1$ MeV and $\langle E_{\nu,x}\rangle = 26.4$ MeV, although that particular simulation also had an overall antineutrino domination over neutrinos. By comparison, the hierarchy of neutrino energies in the DD2 equation of state simulation is less pronounced: $\langle E_{\bar\nu,e} \rangle = 18.2$ MeV and $E_{\nu,x} = 21.9$ MeV.\footnote{The average energies were calculated from the RMS energies given in Ref.~\cite{2016PhRvD..93d4019F}, assuming a neutrino degeneracy parameter $\eta_{\nu,\alpha} = 3$ for all neutrino types.}
	
\begin{table}[ht]
\begin{ruledtabular}
\begin{tabular}{lcdr}
\text{Parameter} & 
\text{Value} \\
\colrule
	$L_{\nu,e}$ & $1.5\times 10^{53}\, \text{erg/s}$\\
   	$L_{\bar\nu,e}, L_{\nu,\bar\nu,x}$ & $2\times 10^{53}\,\text{erg/s}$\\
	$\langle E_{\nu,e}\rangle$ & $11\, \text{MeV}$\\
	$\langle E_{\bar\nu,e}\rangle$ & $18\, \text{MeV}$\\
	$\langle E_{\nu,\bar\nu,x}\rangle$ & $25\, \text{MeV}$\\
	$\delta m^2_\text{atm}$ & $2.4\times 10^{-3}\, \text{eV}^2$\\
	$\theta_\text{V}$ & $8.7^\circ$\\
\end{tabular}
\caption{Parameters used for one of our two-flavor simulations that exhibited the bipolar spectral swap.}
\label{table:params}
\end{ruledtabular}
\end{table}

\begin{figure}[ht]
\centering
\includegraphics[width=.5\textwidth]{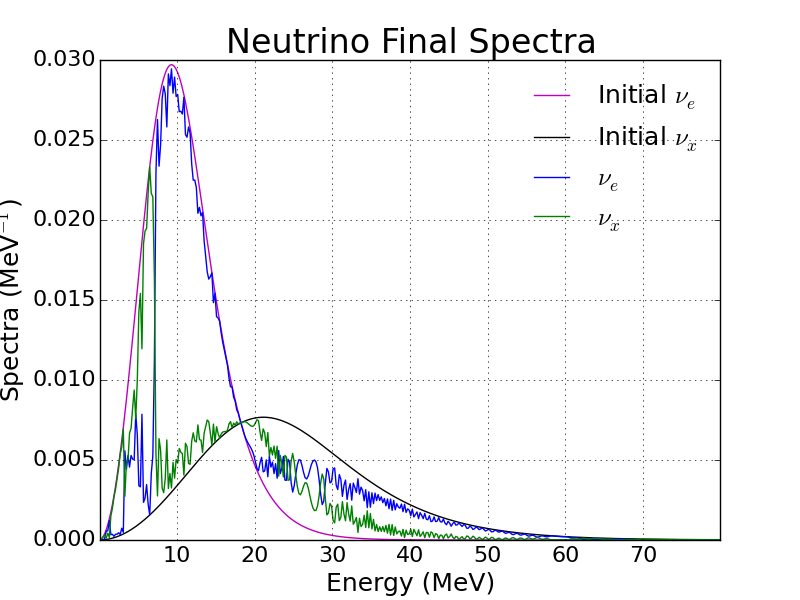}
\caption{Shown here are the initial $\nu_e$ (magenta) and $\nu_x$ (black) energy spectra, as well as the final $\nu_e$ (blue) and $\nu_x$ (green) spectra, at a distance of $5000$ km from the neutrino disk along the polar-axis trajectory, for the simulation with parameters as described in table~\ref{table:params}. A flavor swap is seen to occur at an energy of approximately $8\, \text{MeV}$, and a high energy electron neutrino tail is also seen to develop. This tail of high energy electron neutrinos could potentially affect the electron fraction significantly.} 
\label{fig:diskspectra}
\end{figure}

Figure \ref{fig:diskspectra} shows the initial and final neutrino energy spectra for $\nu_e$ and $\nu_x$ flavors, along the chosen trajectory described earlier. The final spectra were taken from our results at distance of $5000$ km from the neutrino disk, by which point the collective oscillations have stabilized. The spectra were normalized in the same way as in \cite{Duan06a}. The first feature that is readily apparent is a stepwise flavor swap which occurred at a critical energy $E_C\approx 8 \, \text{MeV}$. Electron and $x$-neutrinos with energies below this swap energy mostly converted into each other. This is consistent with previous studies of supernova neutrino flavor evolution in the normal mass hierarchy~\cite{Duan08,Cherry:2010lr}. Separately from the flavor swap at low energies, at energies greater than a threshold energy of about $E_H\approx 20 \, \text{MeV}$, a secondary flavor swap occurred and a significant portion of $x$-neutrinos were converted into electron neutrinos and vice versa. Neutrinos of intermediate energies, i.e., in between the critical and threshold energies, $E_C\lesssim E_\nu \lesssim E_H$, mostly remained in their initial states. As there are many more high energy $x$-neutrinos than electron neutrinos in the initial state, this secondary swap at energies greater than $E_H$ means that a net excess of high energy electron neutrinos develops in the tail as compared to the initial distribution.

\begin{figure}[ht]
\centering
\includegraphics[width=.5\textwidth]{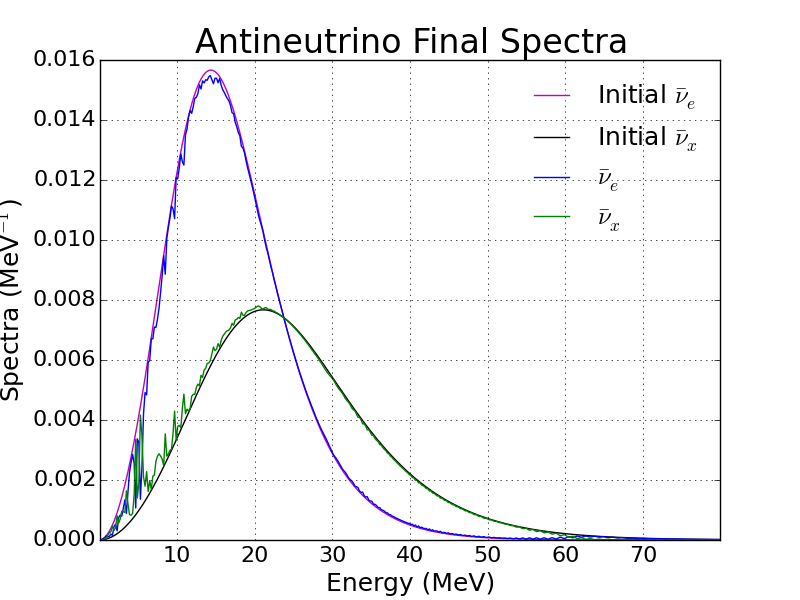}
\caption{Shown here are the initial $\bar\nu_e$ (magenta) and $\bar\nu_x$ (black) energy spectra, as well as the final $\bar\nu_e$ (blue) and $\bar\nu_x$ (green) spectra at a distance of $5000$ km from the neutrino disk, along the polar-axis trajectory, for the same simulation as Fig.~\ref{fig:diskspectra}. Antineutrinos did not significantly convert from one flavor to another.} 
\label{fig:diskspectraanti}
\end{figure}

Figure \ref{fig:diskspectraanti} shows the initial and final energy spectra in the antineutrino sector. As is evident from this figure, a spectral swap occurs in the antineutrino sector at energy $E_C$, though it is not as pronounced as the swap in the neutrino sector. However, the second swap at high energies did not occur in the antineutrino sector. As a result, no high energy electron-antineutrino tail developed in this case. 

\begin{figure*}[ht]
\centering
\subfloat{\includegraphics[width=.5\textwidth]{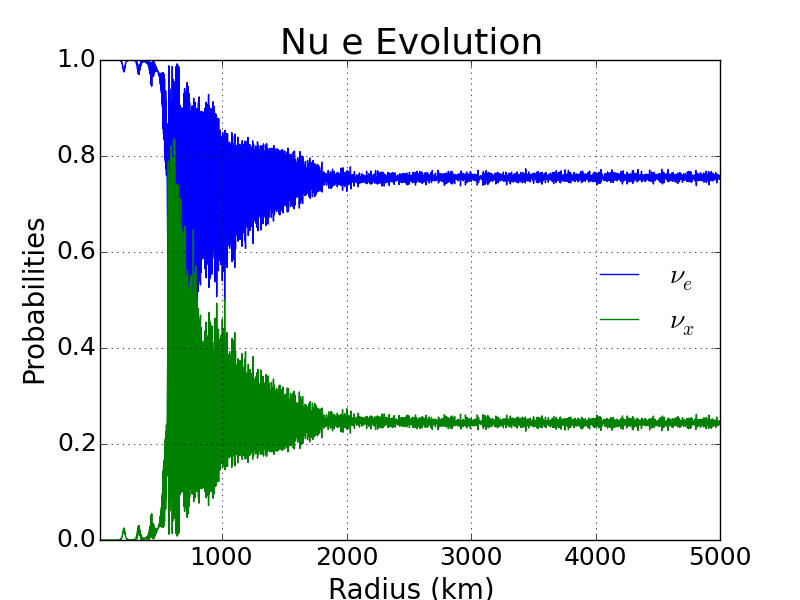}}
\subfloat{\includegraphics[width=.5\textwidth]{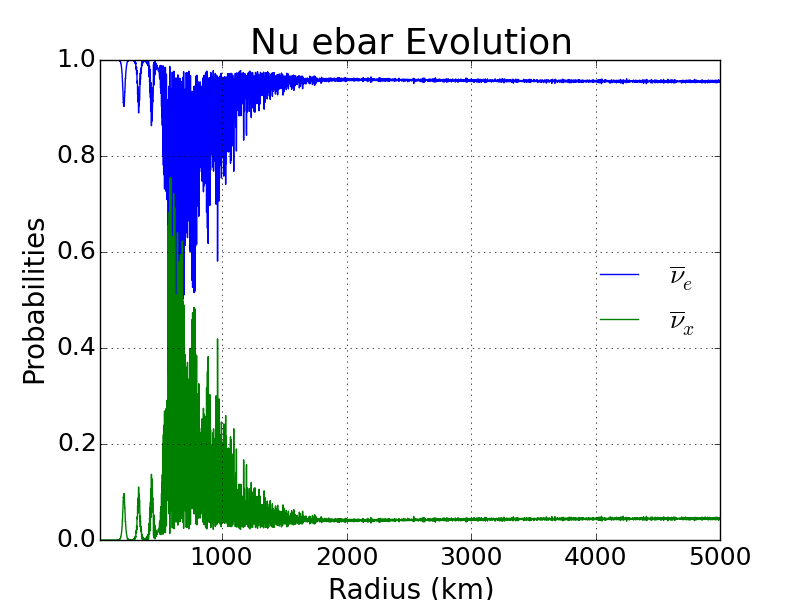}}
\caption{These plots show the energy-averaged neutrino flavor evolution for a neutrino (left) and an antineutrino (right) initially in the electron flavor state, for the simulation with the parameters listed in table~\ref{table:params}. We can see that significant flavor transformations begin to occur at a radius of approximately $600 \, \text{km}$ and these flavor oscillations stabilize at a radius of approximately $2000 \, \text{km}$}  
\label{fig:NueEvolution}
\end{figure*}

The plots in Fig.~\ref{fig:NueEvolution} show the energy-averaged probabilities for a (anti-)neutrino that started out in the electron flavor state to be in the electron or $x$ flavor states as a function of distance. It is evident from these figures that, in both the neutrino and antineutrino sectors, collective flavor evolution phenomena set in at a radius of approximately $500\text{--}600 \, \text{km}$. Minimal flavor transformation occurred closer to the neutrino disk; however, significant large scale flavor conversion does not set in until farther out. The neutrino flavors oscillate rapidly with distance for approximately $1500 \, \text{km}$ and then stabilize around the final values at a radius of approximately $2000 \, \text{km}$.

\begin{figure*}[ht]
\centering
\subfloat{\includegraphics[width=.5\textwidth]{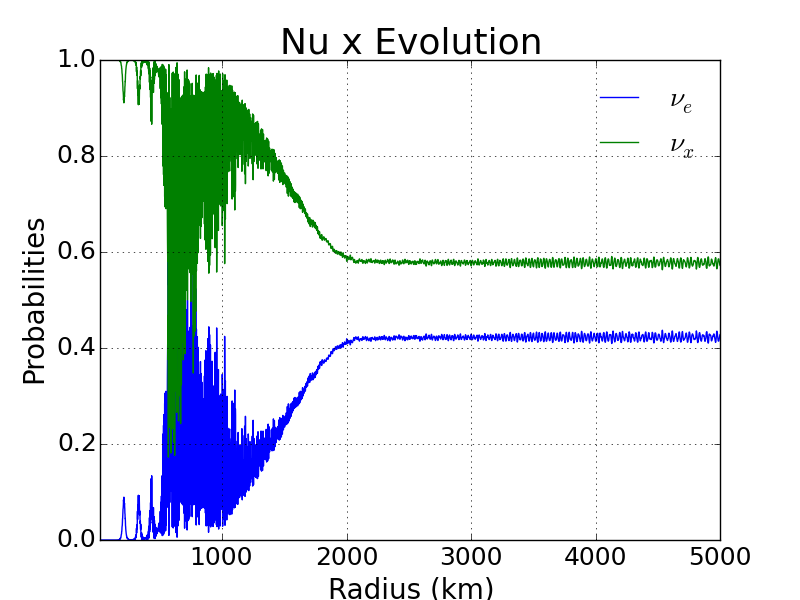}}
\subfloat{\includegraphics[width=.5\textwidth]{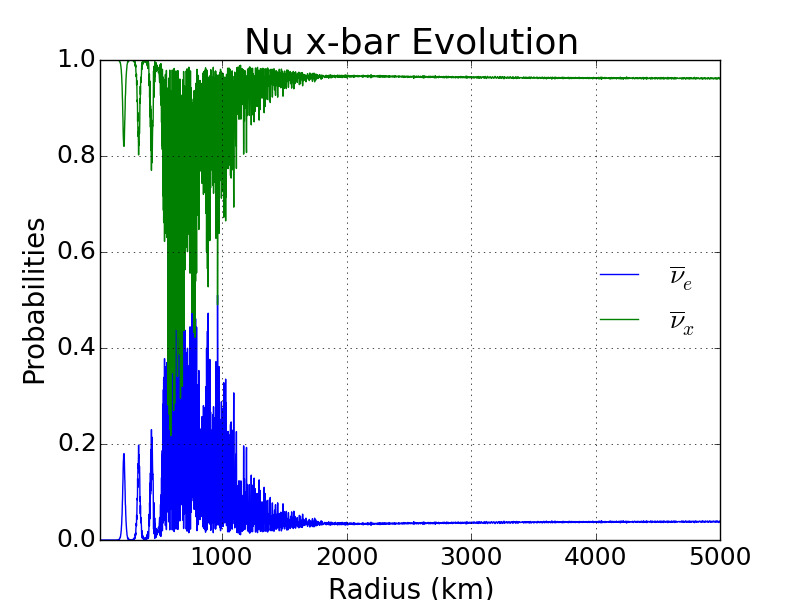}}
\caption{This is the energy-averaged evolution of neutrino flavors for a neutrino (left) and antineutrino (right) initially in the $x$-flavor state.} 
\label{fig:NumuEvolution}
\end{figure*}

The plots in Fig.~\ref{fig:NumuEvolution} show the energy-averaged probabilities for a (anti-)neutrino that started out in the $x$ flavor state to be in the electron or $x$ flavor states as a function of distance. These plots and the two previous plots demonstrate that although both the neutrino and antineutrino sectors go through rapid flavor oscillations, the antineutrino sector did not sustain significant overall flavor transformation while the neutrino sector did. As much as 40\% of $x$-neutrinos were converted into electron neutrinos after the oscillations stabilized, while only a very small percentage of anti-$x$-neutrinos were converted into anti-electron-neutrinos. Most interestingly, as can be seen from the neutrino flavor evolution plots, while approximately 40\% of initial $x$-neutrinos converted into electron neutrinos, only approximately 20\% of initial electron neutrinos converted into $x$-neutrinos. Considering that the total initial luminosity of $x$-neutrinos was higher than that for electron neutrinos, and noting that preferentially higher energy $\nu_x$ were converted to $\nu_e$, while lower energy $\nu_e$ were converted to $\nu_x$ (see Fig.~\ref{fig:diskspectra}), it can be concluded that there is a net excess of electron neutrino energy flux resulting from this transformation. This phenomenon could potentially have a negative effect on the neutron excess, and thereby, on the prospects for $r$-process nucleosynthesis in ejecta moving out along this direction.

\begin{figure*}[ht]
\centering
\subfloat{\includegraphics[width=.5\textwidth]{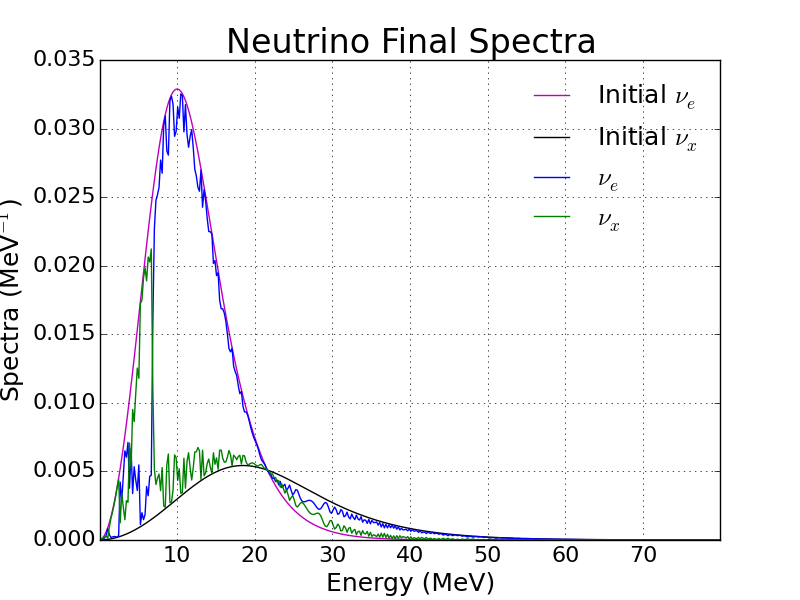}}
\subfloat{\includegraphics[width=.5\textwidth]{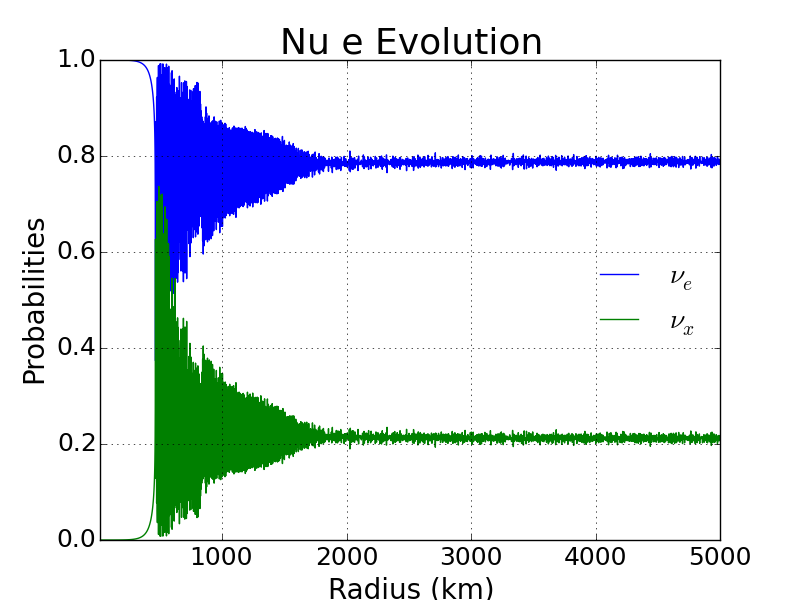}}
\caption{Results for a calculation run using the luminosities and average energies adopted from the DD2 equation of state simulation in Ref.~\cite{2016PhRvD..93d4019F}. This calculation also demonstrates a bipolar spectral swap, qualitatively very similar to the one shown in Fig.~\ref{fig:diskspectra}. (Left) Initial and final ($r = 5000$ km) neutrino energy spectra. (Right) Energy-averaged flavor evolution of a neutrino initially emitted in the electron flavor state.} 
\label{fig:DD2evol}
\end{figure*}

For completeness, we have also included an abbreviated set of plots (Fig.~\ref{fig:DD2evol}) showing the results of a flavor evolution calculation using the exact luminosities and spectra from the DD2 equation of state simulation in Ref.~\cite{2016PhRvD..93d4019F} (Table III from this reference). The left panel shows the initial and final ($r = 5000$ km) spectra for $\nu_e$ and $\nu_x$, whereas the right panel shows the energy-averaged flavor evolution probabilities for a neutrino initially in the electron flavor state. Qualitatively, these results can be seen to be almost identical to the corresponding plots from Figs.~\ref{fig:diskspectra} and \ref{fig:NueEvolution}, even though this particular set of initial conditions does not exhibit as strong of an energy hierarchy between $\bar\nu_e$ and $\nu_x$ as the parameters in Table~\ref{table:params}, as discussed earlier in this section.

\subsection{Three Flavor Bipolar Swap Results}\label{sec:3fbipolarresult}

Table \ref{table:3fparams} shows the parameters we used in our three-flavor oscillation simulations. Luminosities in the $x$-neutrino sector used in our two-flavor simulations (Table \ref{table:params}) were split evenly among the $\mu$ and $\tau$ flavors in three-flavor simulations in order to keep constant the total neutrino luminosity among all flavors. Three-flavor neutrino mixing will involve both the atmospheric neutrino mass-squared splitting $\delta m^2_\text{atm}$ and the solar neutrino mass-squared splitting $\delta m^2_\odot$ as well as three total mixing angles $\theta_{12},\theta_{13},\theta_{23}$ and an as yet unknown $CP$-violating phase $\delta_{CP}$. In our calculations, the $CP$-violating phase was set to 0.  

\begin{table}[tb]
\begin{ruledtabular}
\begin{tabular}{lcdr}
\text{Parameter} & 
\text{Value} \\
\colrule
	$L_{\nu,e}$ & $1.5\times 10^{53}\, \text{erg/s}$\\
	$L_{\bar\nu,e}$ & $2\times 10^{53}\, \text{erg/s}$\\
   	$L_{\nu,\bar\nu,\mu,\tau}$ & $1\times 10^{53}\,\text{erg/s}$\\
	$\langle E_{\nu,e}\rangle$ & $11\, \text{MeV}$\\
	$\langle E_{\bar\nu,e}\rangle$ & $18\, \text{MeV}$\\
	$\langle E_{\nu,\bar\nu,\mu,\tau}\rangle$ & $25\, \text{MeV}$\\
	$\delta m^2_\text{atm}$ & $2.4\times 10^{-3}\, \text{eV}^2$\\
	$\delta m^2_\odot$ & $7.6\times 10^{-5}\,\text{eV}^2$\\
	$\theta_\text{12}$ & $34.4^\circ$\\
	$\theta_\text{13}$ & $8.7^\circ$\\
	$\theta_\text{23}$ & $45^\circ$\\
	$\delta_\text{CP}$ & $0^\circ$\\
\end{tabular}
\caption{Parameters used our three flavor simulation. Between the two flavor and three flavor case, the $x$-neutrino luminosity was split equally among the $\mu$ and $\tau$ flavors.}
\label{table:3fparams}
\end{ruledtabular}
\end{table}

\begin{figure}[ht]
\centering
\includegraphics[width=.5\textwidth]{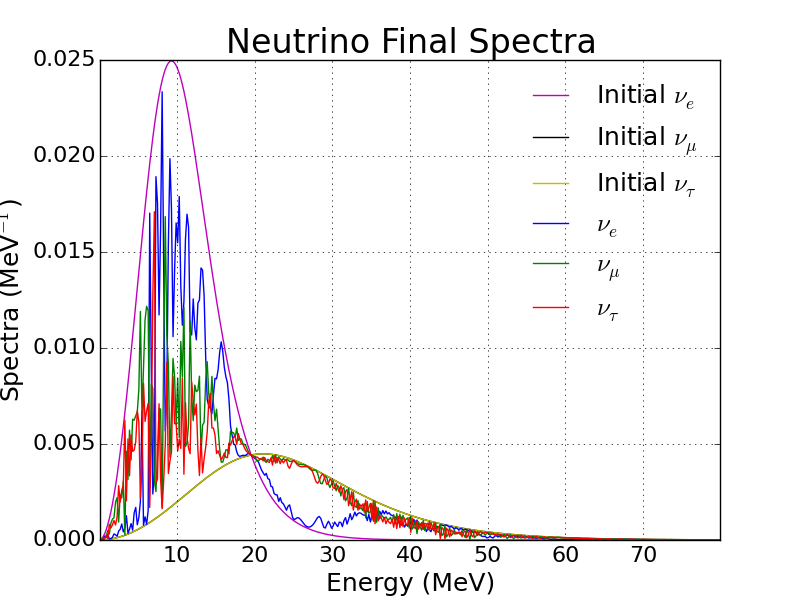}
\caption{This is the final energy distribution spectra of neutrinos for a three flavor simulation. Initial $\mu$ and $\tau$ neutrinos have the same energy spectra and so overlap on this graph. The red line represents both flavors.} 
\label{fig:FinalSpectradisk3f}
\end{figure}

Figure \ref{fig:FinalSpectradisk3f} shows the final energy distribution spectra for the electron neutrino, muon neutrino, and tau neutrino in our three flavor simulation. As in the two-flavor simulation, a high-energy electron neutrino tail develops in this case. However, because of the presence of possible transformations into a third flavor, the high energy electron neutrino tail is less pronounced than in the two-flavor cases, particularly in the energy range of roughly $20\text{--}30 \, \text{MeV}$. Moreover, at energies of approximately $8\text{--}20 \, \text{MeV}$, the electron neutrinos significantly transform into other flavors of neutrinos, which was not the case in the two-flavor simulations. As a result, three-flavor simulations indicate fewer total electron neutrinos present at large distance than do two-flavor simulations.  

\begin{figure}[ht]
\centering
\includegraphics[width=.5\textwidth]{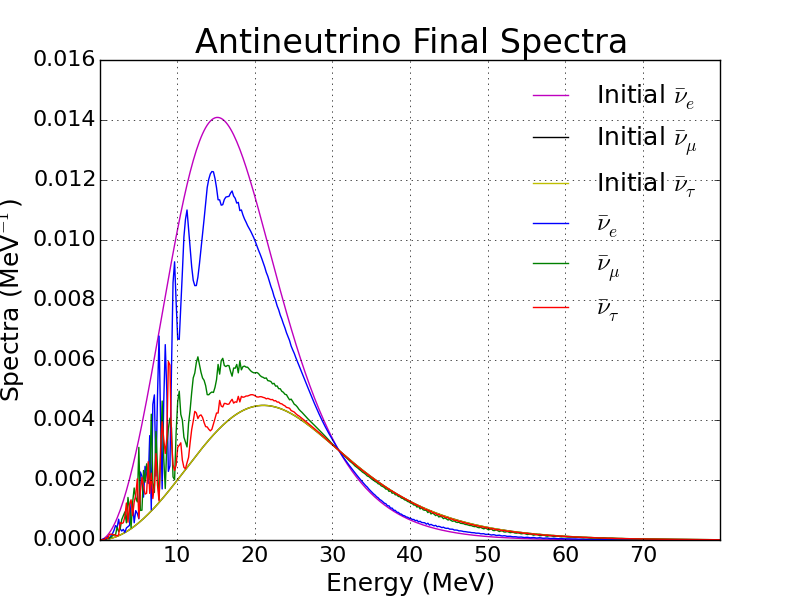}
\caption{Same as Fig.~\ref{fig:FinalSpectradisk3f}, but for anti-neutrinos.} 
\label{fig:FinalSpectradisk3fanti}
\end{figure}

Figure \ref{fig:FinalSpectradisk3fanti} shows the final energy distribution spectra for the antineutrino sector in our three-flavor simulations. Relatively more collective flavor conversion occurred in the antineutrino sector for three flavor simulations than for two flavor simulations. However, the flavor transformation in the antineutrino sector is nevertheless not as significant as that in the neutrino sector. Specifically, no high-energy electron antineutrino tail develops.

\begin{figure*}[ht]
\centering
\subfloat{\includegraphics[width=.5\linewidth]{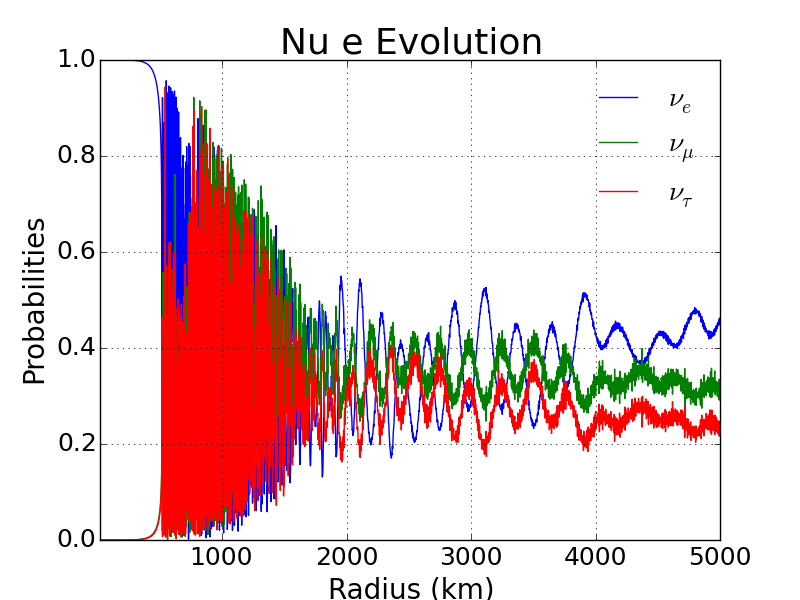}}
\subfloat{\includegraphics[width=.5\linewidth]{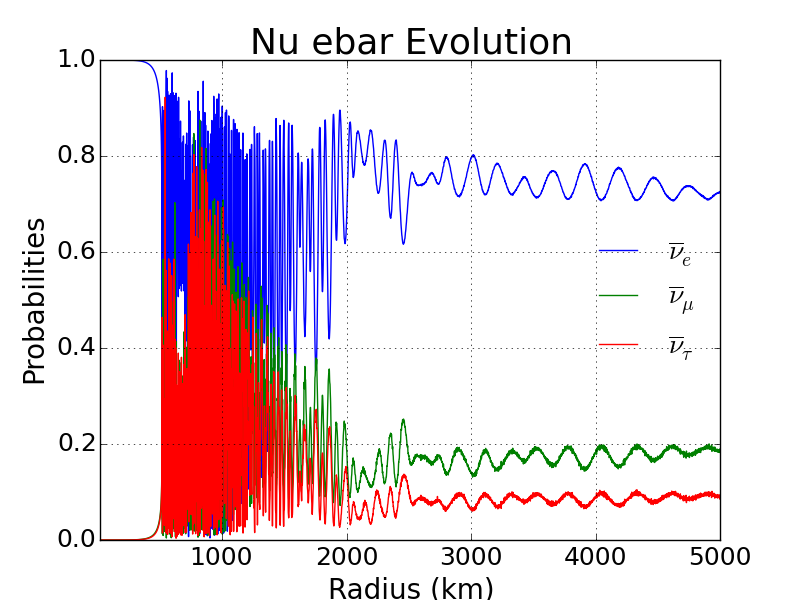}}\\
\subfloat{\includegraphics[width=.5\linewidth]{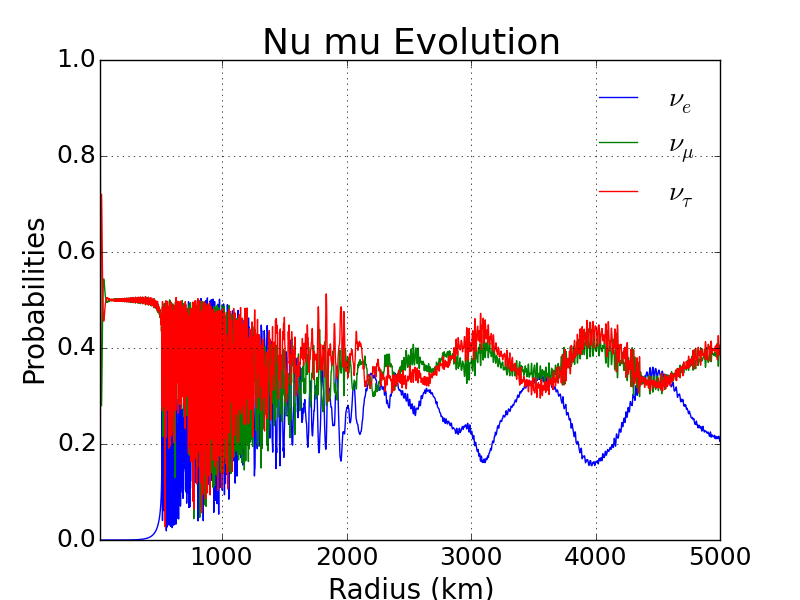}}
\subfloat{\includegraphics[width=.5\linewidth]{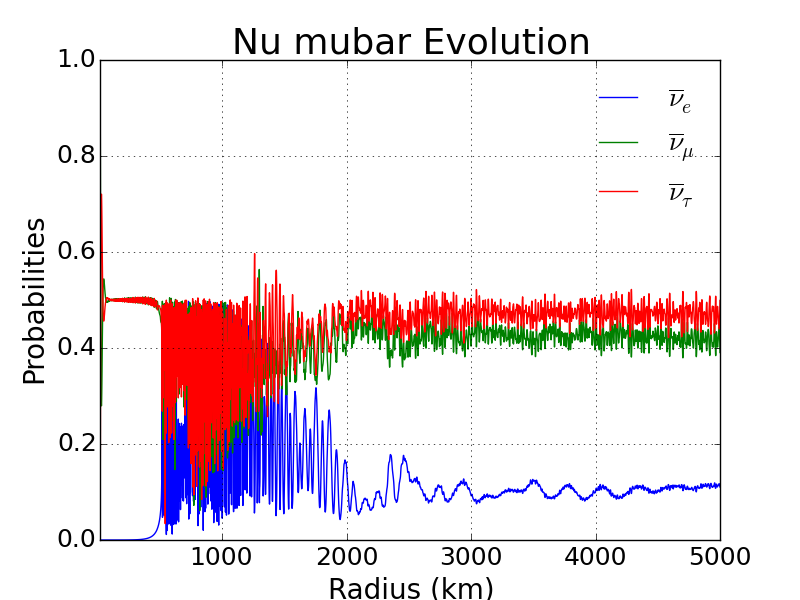}}\\
\subfloat{\includegraphics[width=.5\linewidth]{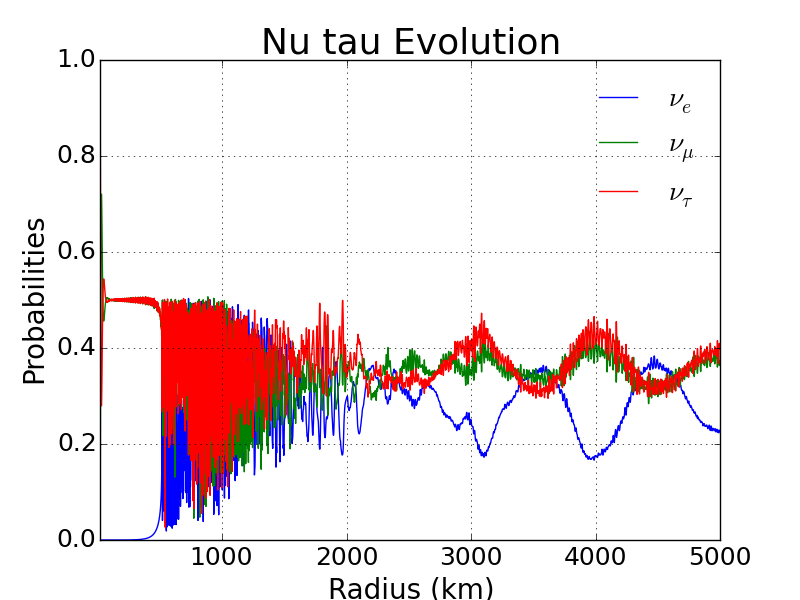}}
\subfloat{\includegraphics[width=.5\linewidth]{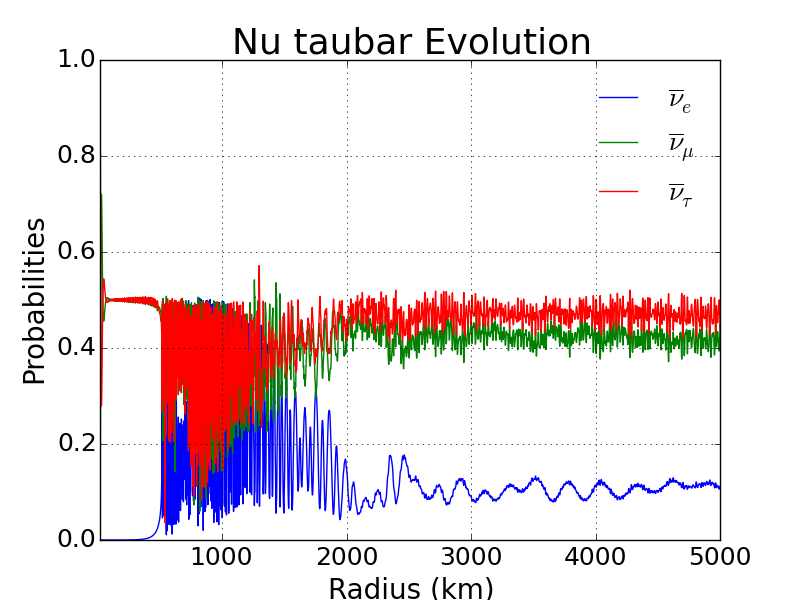}}
\caption{Plots of the energy-averaged flavor evolution of neutrinos which start in the various initial states. We can see here that still interesting neutrino flavor transformations seem to occur at a radius of approximately $600 \, \text{km}$. However, here, the neutrino flavor transformations do not seem to stabilize as much as in the two flavor case. Although the antineutrinos mix more in the three flavor case than the two-flavor case, it still does not convert as many $\mu$ and $\tau$ flavor antineutrinos into the electron flavor as in the neutrino sector.}
\label{fig:3flavorevolution}
\end{figure*}

Figure \ref{fig:3flavorevolution} shows all the plots for the energy-averaged neutrino probability evolution in a three-flavor simulation. The collective neutrino transformation begins at a radius of $600 \, \text{km}$. This is same as in the two-flavor case. The mu and tau neutrinos and antineutrinos remain nearly maximally mixed throughout the simulation. Since these neutrinos are nearly maximally mixed in vacuum, and since they experience nearly identical interactions in medium, they evolve nearly identically in our simulations.

\subsection{Low Luminosity and Low Density Results}\label{lowlumlowdens}

As mentioned before, neutron star merger environments can manifest a multitude of different initial conditions. Different simulations using different equations of state, or initial configurations of the neutron stars, produce different density profiles, neutrino luminosities, and neutrino spectra. In order to explore alternative initial conditions, we ran a simulation where the initial baryon mass density at the neutrino disk was lowered from $n_{b,0} = 10^8 \, \text{g/cm}^3$ to $n_{b,0} = 2.5\times 10^{6}\, \text{g/cm}^3$. This density is closer to the initial density found in the simulations in Ref.~\cite{2016arXiv160705938F}. The neutrino spectral shape parameters ($\langle E_{\nu,\alpha} \rangle$ and $\eta_{\nu,\alpha}$) were kept the same as those used in Sec.~\ref{sec:2fbipolarresult}, but the luminosities were lowered by an order of magnitude across the board: $L_{\nu,e} = 1.5 \times 10^{52} \, \text{erg/s}$, $L_{\nu,\bar{e},x,\bar{x}} = 2 \times 10^{52} \, \text{erg/s}$, closer to the luminosities found in Ref.~\cite{2016arXiv160705938F}.
 
\begin{figure*}[ht]
\centering
\subfloat{\includegraphics[width=.5\textwidth]{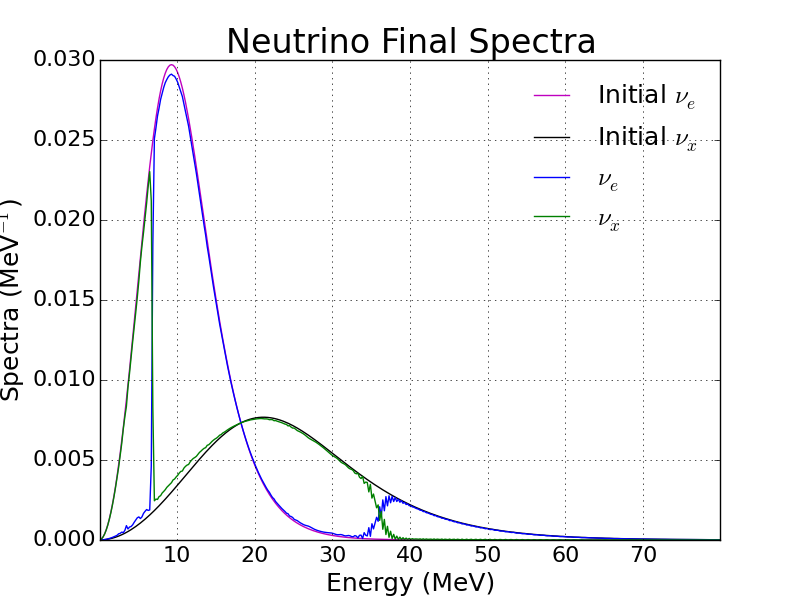}}
\subfloat{\includegraphics[width=.5\textwidth]{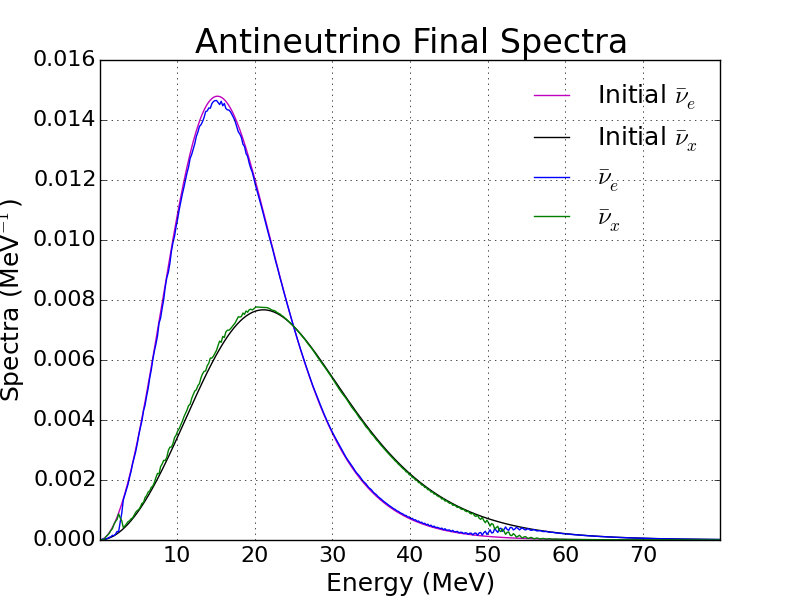}}
\caption{These figures show the final energy spectra of neutrinos and antineutrinos, respectively, run with lowered luminosity and density conditions from those simulations shown in Secs.~\ref{sec:2fbipolarresult} and \ref{sec:3fbipolarresult}. In the neutrino sector, we can see a very clear bipolar swap at an energy of $\approx 8 \, \text{MeV}$. This is perhaps the clearest bipolar swap result found in our simulations. In addition, the high energy electron neutrino tail for neutrinos of energy $\gtrsim 38 \, \text{MeV}$ is very pronounced.} 
\label{fig:spectralowlumlowdens}
\end{figure*}

Figure \ref{fig:spectralowlumlowdens} shows the final spectra for neutrino and antineutrinos. This result is the clearest example of a bipolar spectral swap that was found in our simulations. There is a very sharp swap at low energies $E_C \approx 8 \, \text{MeV}$. In addition, the high energy electron neutrino tail is still present. However, the energy at which the tail manifests is higher, $\approx 38$ MeV. The transition to the high energy electron neutrino tail also is much more pronounced and sharp than in the simulations described in Secs.~\ref{sec:2fbipolarresult} and \ref{sec:3fbipolarresult}. Likewise, in the antineutrino sector, there are analogous, although less pronounced, effects of spectral swaps at low energies $\lesssim 4 \, \text{MeV}$ and high energies $\gtrsim 51 \, \text{MeV}$. All spectral features in both the neutrino and antineutrino sectors are more pronounced and sharper in this simulation compared to the ones in Secs.~\ref{sec:2fbipolarresult} and \ref{sec:3fbipolarresult}.      

\begin{figure*}[ht]
\centering
\subfloat{\includegraphics[width=.5\textwidth]{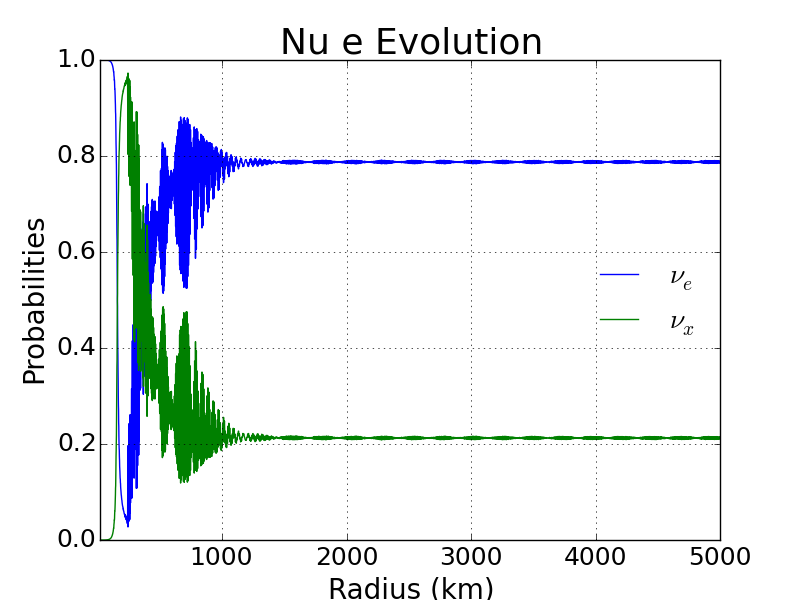}}
\subfloat{\includegraphics[width=.5\textwidth]{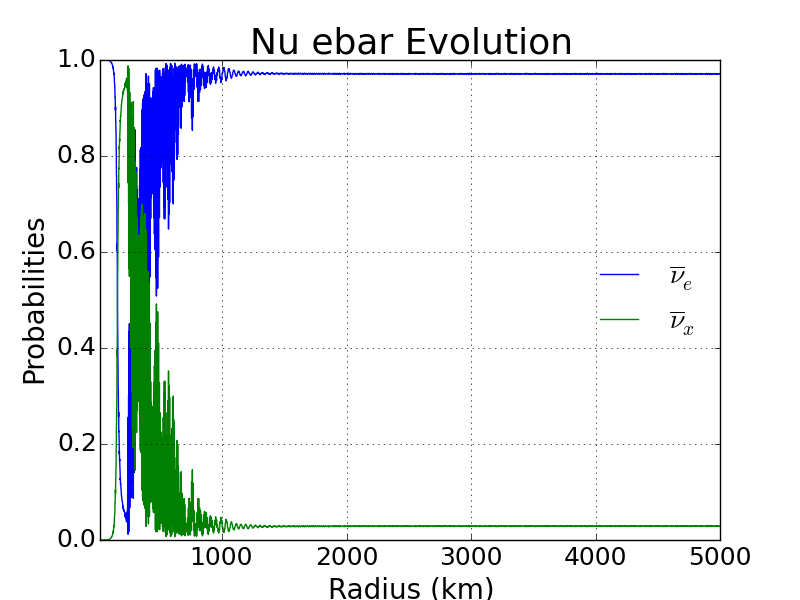}}
\caption{These figures show the energy-averaged flavor evolution of a neutrino and antineutrino, respectively, which started out in the electron flavor state for a simulation with a lowered neutrino luminosity and initial density. We can see that flavor evolution sets in much closer to the neutrino disk than the simulations shown with a higher luminosity and density in Secs.~\ref{sec:2fbipolarresult} and \ref{sec:3fbipolarresult}. Here the flavor evolution begins around a radius of $\approx 100 \, \text{km}$.} 
\label{fig:NueEvolutionlowlumlowdens}
\end{figure*}

\begin{figure*}[ht]
\centering
\subfloat{\includegraphics[width=.5\textwidth]{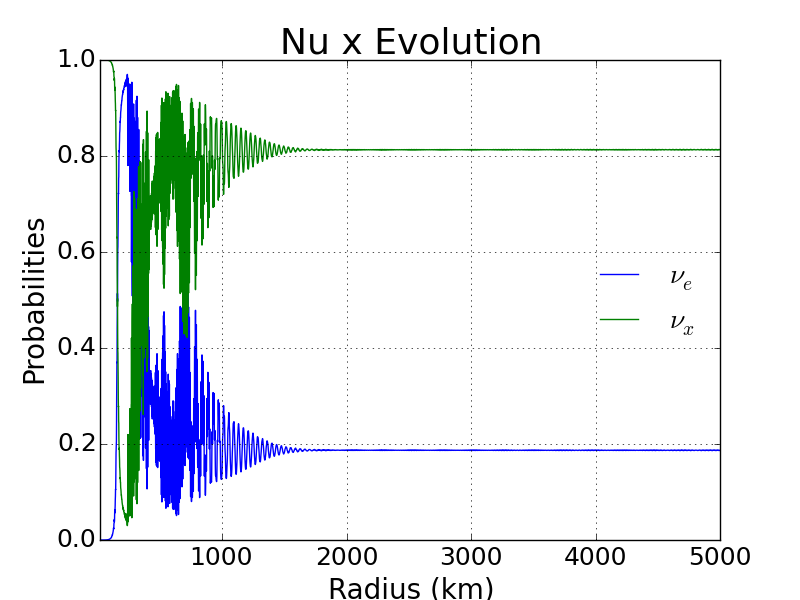}}
\subfloat{\includegraphics[width=.5\textwidth]{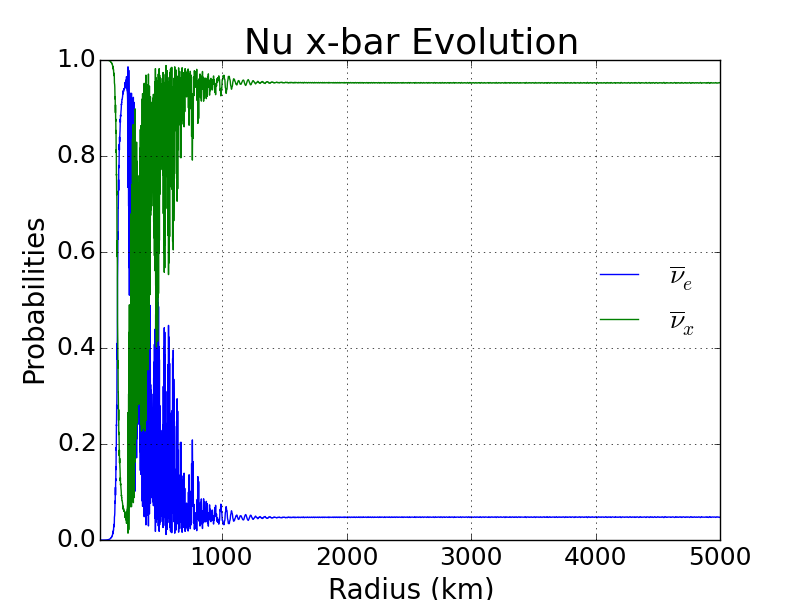}}
\caption{This is the evolution of neutrino flavors for a neutrino and antineutrino initially in the x flavor state in a simulation with a lowered luminosity and initial density.} 
\label{fig:NumuEvolutionlowlumlowdens}
\end{figure*}

Figures \ref{fig:NueEvolutionlowlumlowdens} and \ref{fig:NumuEvolutionlowlumlowdens} show the energy-averaged neutrino and antineutrino flavor evolution for an initially electron flavor and initially $x$ flavor neutrino, respectively. These reveal that significant neutrino flavor evolution begins much closer to the neutrino disk, at a radius of $\approx 100 \, \text{km}$, than does the analogous flavor transformation in the simulations discussed in Secs.~\ref{sec:2fbipolarresult} and \ref{sec:3fbipolarresult}. As we will discuss in Sec.~\ref{sec:discussion}, this is to be expected. It is also easier and clearer to see here that the neutrino flavor evolution begins as synchronized oscillations, before eventually settling down into a bipolar spectral swap.

\section{Discussion}\label{sec:discussion}

Collective neutrino oscillations driven by the nonlinear aspects of $H_{\nu\nu}$ can occur in both the core-collapse supernova and BNS merger environments. This is unsurprising at some level, because both these astrophysical venues are characterized by prodigious neutrino fluxes. A particularly interesting collective neutrino flavor oscillation feature, the bipolar spectral swap, can appear in both environments as well. Ref.~\cite{2016arXiv160705938F} showed that bipolar collective oscillations can occur in the BNS merger environment along oblique trajectories between the polar axis and the neutrino disk, in the inverted neutrino mass hierarchy and in antineutrino dominated conditions. The calculations presented here show that bipolar collective oscillations, along with ensuing spectral swaps, can also occur in neutrino-dominated conditions in the normal mass hierarchy. This can have potentially interesting implications, as we will discuss in Sec.~\ref{sec:ye}.

\subsection{Flavor Evolution}

As discussed above, our simulations with the normal neutrino mass hierarchy show bipolar collective flavor oscillations which produce spectral swaps.  In fact, we find a two-tiered stepwise spectral swap which gives rise to not only the usual swapping of electron and $x$-neutrinos at low energies, but also a secondary partial swap of flavors at high energies, resulting in an enhanced high energy electron neutrino tail. At low radii ($r \lesssim 100 \,$km for the low density and low luminosity simulations, and $r \lesssim 500\,$km for the high density and high luminosity simulations), the large matter and neutrino potentials keep the neutrinos mostly locked in their initial flavor states. This is a consequence of the instantaneous in-medium mass eigenstates being driven apart, thereby suppressing the corresponding in-medium flavor mixing parameters. At intermediate radii ($r \approx 100\text{--}500\,$km and $r \approx 500\text{--}800\,$km for the aforementioned two cases), the neutrinos undergo synchronized oscillations. Eventually (at $r \gtrsim 500\,$km and $r > 800\,$km) bipolar spectral swaps develop\footnote{For a helpful illustration of these phenomena, please refer to the movies of neutrino spectra as a function of radius, which we have uploaded here~\cite{movies}. The movies titled \lq\lq Neutrino Spectra Two Flavor Bipolar Swap\rq\rq\ and \lq\lq Neutrino Spectra Low Luminosity Low Density\rq\rq\ show the flavor evolution along the polar axis in the high luminosity, high density (Sec.~\ref{sec:2fbipolarresult}), and the low luminosity/density (Sec.~\ref{lowlumlowdens}) cases, respectively.}.

Qualitatively speaking, this is similar to the behavior exhibited in flavor transformation simulations in core-collapse supernova environments. For instance, the final spectrum that we see in our simulations looks qualitatively similar to the final spectrum presented for the normal neutrino mass hierarchy in \cite{Duan06a} (see Fig. 7 (a) in that paper). As such, it is likely that the same physical phenomenon which guided the flavor evolution in the simulations presented in that paper guides the flavor evolution here. The neutrino spectra that we used were quite similar to those in Ref. \cite{Duan06a}; however, it bears noting that the characteristic radii at which collective neutrino transformations begin and end are quite different between our simulations discussed in Secs.~\ref{sec:2fbipolarresult} and \ref{sec:3fbipolarresult}, and those in Ref.~\cite{Duan06a}. This results from the much higher baryon density (by about two orders of magnitude) that we used in these simulations, as compared to the supernova environment analyzed in Ref.~\cite{Duan06a}, as well as the high luminosity. If the luminosity and density are lowered, significant neutrino flavor conversion does occur closer to the neutrino disk, as observed in our low luminosity and low density simulation.

Refs.~\cite{Duan06a, PhysRevD.76.125008, PhysRevD.77.113007, Raffelt07} demonstrate how a geometric picture of flavor evolution can be developed in the two-flavor case by mapping the neutrino modes (described as $SU(2)$ spinors) to their equivalent $SO(3)$ representations, termed either \lq\lq Neutrino flavor iso-spins (NFIS)\rq\rq\ or \lq\lq Polarization vectors\rq\rq. As shown in these references, this can be used to explain the bipolar spectral swap at low energies $E_\nu \lesssim E_C$ using an analytic analysis of neutrino flavor evolution. However, at neutrino energies above the threshold energy $E_H$, the neutrinos may not be locked into the collective bipolar modes. These neutrinos may be converted via a background-assisted MSW mechanism to form the high energy electron neutrino tail that we see in both the high luminosity, high density and low luminosity, low density simulations.

The major difference between the physical conditions employed in our simulations and those in supernova simulations of neutrino flavor evolution is the geometric dilution of neutrinos in the two venues, i.e., a spherical neutrino source in the supernova case versus a disk-like source in the BNS merger case. Differences in neutrino luminosity and baryon number density mostly serve to change the relative locations of the onset of collective neutrino flavor evolution. The mechanisms through which the neutrino flavors transform, however, are not sensitive to this difference in geometric dilution. The bipolar flavor swap requires only that (1) the neutrino Hamiltonian dominate at some point to bring the neutrinos into a synchronized oscillation mode, and (2) the neutrino Hamiltonian must then gradually decrease with increasing radius in order for the flavor conversion to remain in the adiabatic regime. If these conditions are met, the bipolar spectral swap phenomenon seen in results of our simulations is robust to the details of the neutrino geometric dilution.

\subsection{Electron Fraction Ramifications}\label{sec:ye}

A potentially important question is whether the collective oscillation induced modification of the neutrino and antineutrino energy spectra could affect the material composition, i.e., the electron fraction $Y_e\equiv n_e/n_b$, defined as the ratio of the net electron number density to the baryon number density. The electron fraction can be important in determining $r$-process yields, being directly related to the neutron-to-proton ratio, $n/p = (1/Y_e) - 1$. If we follow a fluid element as it leaves the merger environment, the local electron fraction would be determined by the interplay between the neutron-proton interconversion, via the weak capture processes of Eq. (\ref{eqn:reactions}), and the matter outflow rate. Here we assume that the matter consists of free nucleons, a fair assumption given the likely high entropy in this region of binary neutron star merger outflow. However, if the matter in the region of interest has lower entropy, and therefore potentially a lower free nucleon fraction and higher nuclear mass fraction, then the neutrino spectral changes we discuss may not have as large an effect on the electron fraction and, consequently, the requisite conditions for the r-process. This is because nucleons locked up in heavy nuclei have generally lower available weak interaction strength than do free neutrons and protons. 

We can label the rates (in units $s^{-1}$) for the reactions in Eq. (\ref{eqn:reactions}) as $\lambda_{\nu_e n}$, $\lambda_{e^-p}$, $\lambda_{\bar{\nu}_e p}$, and $\lambda_{e^+ n}$ where the subscripts refer to particles \textit{entering} each reaction. Two of these rates destroy neutrons and the other two produce neutrons. The electron fraction then evolves according to these rates, as a competition between neutron production and destruction (where $Y_p$ and $Y_n$ are the proton and neutron fractions respectively):
\begin{equation}
\frac{d}{dt}Y_e=(\lambda_{\nu_e n}+\lambda_{e^+ n})Y_n - (\lambda_{\bar{\nu}_e p}+\lambda_{e^-p})Y_p \, .
\label{eqn:Ye}
\end{equation}

Imposing charge neutrality, and assuming for our purposes that baryons are composed purely of neutrons and protons, we have $Y_p=Y_e$, and $Y_n=1 - Y_e$. We can then turn Eq. (\ref{eqn:Ye}) into an ordinary differential equation in $Y_e$:
\begin{equation}
\frac{d}{dt}Y_e = \lambda_1 - \lambda_2\,Y_e,
\label{eqn:YeEvol}
\end{equation}
where we have defined $\lambda_1\equiv \lambda_{\nu_e n}+\lambda_{e^+ n}$ and $\lambda_2\equiv\lambda_{\nu_e n}+\lambda_{e^-p}+\lambda_{\bar{\nu}_e p}+\lambda_{e^+ n}$~\cite{Qian93}.

Knowing the weak interaction rate and outflow velocity history for a given fluid element will then tell us about the way it evolves in $Y_e$. Since our interest lies in evaluating the effects of neutrino flavor transformations on the weak processes, the rates we have chosen to focus on in what follows are the rates of neutron destruction and production via neutrino capture processes, $\lambda_{\nu_en}$ and $\lambda_{\bar{\nu}_ep} $. These rates depend on the neutrino and antineutrino fluxes and distribution functions, and on the interaction cross-sections. Generically, dropping the subscripts so that the quantities may represent either of the two processes, these rates can be expressed as:
\begin{equation}\label{eqn:lambda}
\lambda(r) = \int_{\bm{p}} \sigma(\bm{p})\, dn_\nu(\bm{p},r) \, .
\end{equation}

Here $\sigma$ is the appropriate neutrino interaction cross-section, and $dn_\nu = dn_{\nu_e}$ or $dn_{\bar\nu_e}$ is the differential number flux of $\nu_e$ or $\bar\nu_e$ at the interaction location. These number fluxes can be expressed in terms of $dn_{\nu,\alpha} (E_\nu)$, i.e., the number density of neutrinos at energy $E_\nu$ with \textit{initial} flavor $\alpha$ (discussed in Sec.~\ref{sec:Hamiltonian}), and the energy-dependent flavor conversion/survival probabilities $P_{\alpha e}(r,E_\nu)$, as follows: 
\begin{equation}
dn_{\nu_e}(r,E_\nu)= \sum_\alpha dn_{\nu,\alpha}(E_\nu)\, P_{\alpha e}(r,E_\nu),
\end{equation}
and similarly for $\bar\nu_e$. Note that, within the single-angle approximation, we can replace the neutrino momentum labels $\bm{p}$ with just the energy $E_\nu$, since the neutrino fluxes/distributions are taken to be independent of emission trajectory.

In particular, to ascertain the conditions (e.g., outflow speeds) that may be required in order to preserve the neutron excess, we shall estimate the neutrino capture rate $\lambda_{\nu_en}$ for some of our simulated environments. The rationale behind choosing to focus on $\lambda_{\nu_en}$ is that one expects the material surrounding the BNS merger disk to be neutron rich to begin with (i.e., $Y_n > Y_p$), making the rate $\lambda_{\nu_en}$ more important in the rate equation [Eq. (\ref{eqn:Ye})] compared to $\lambda_{\bar\nu_ep}$. Moreover, $\nu_e$ capture on $n$ does not have a threshold, unlike $\bar\nu_e$ on $p$, although the effect of this threshold is small at the typical energies in these environments. Borrowing the expression for $dn_{\nu,\alpha}$ from Eq. (\ref{eq:dn}), and integrating over angles, we can write the expression for the rate $\lambda_{\nu_en}$ as
\begin{widetext}
\begin{equation}\label{eqn:lambdanue}
\lambda_{\nu_en}(r) = \int\displaylimits_0^\infty \sum_\alpha \frac{L_{\nu,\alpha}}{2\pi^2 R^2_\nu\langle E_{\nu,\alpha}\rangle}\tilde{f}_{\nu,\alpha}(E_\nu)P_{\alpha e}(r,E_\nu)\cdot 2\pi \left(1 - \frac{r}{\sqrt{r^2+R_\nu^2}}\right) \,\sigma(E_\nu)\, dE_\nu.
\end{equation}
\end{widetext}

The appropriate neutrino capture cross-section in the low momentum-transfer limit is given by~\cite{Dicus:1982bz, Fuller:1995qy, 1980ApJS...42..447F, 1982ApJ...252..715F, 1982ApJS...48..279F, 1985ApJ...293....1F}:
\begin{equation}
\begin{split}
\sigma(E_\nu)&=\frac{2\pi^2(\hbar c)^3}{c}\frac{\ln 2}{\langle ft\rangle}\frac{\langle G\rangle}{(m_e c^2)^5}(E_\nu+Q)^2 \\
			&\equiv\mathcal{C}\,(E_\nu+Q)^2,
\end{split}
\label{eqn:sigma}
\end{equation} 
where $\langle G\rangle$ is the average Coulomb correction factor, $\langle ft\rangle$ contains the pertinent (scattering) matrix elements, and $Q = (m_i - m_f) c^2$ is the $Q$-value of the reaction, i.e., the net rest-mass energy differential between the initial and final constituents. Using $\langle ft \rangle = 10^{3.035}\,$s and $\langle G \rangle = 1$, one can calculate the pre-factor $\mathcal{C}$ to be approximately $9.3 \times 10^{-44}\,\text{cm}^2/\text{MeV}^2$. Note that we have explicitly written all the $c$ and $\hbar$ symbols in Eq. (\ref{eqn:sigma}) to facilitate calculating the cross-section in cm$^2$, rather than in energy units.

For simplicity, we assume here that the constituents of the charged-current neutrino capture processes are the proton, neutron, the electron (or positron), and the nearly massless neutrino (i.e., no heavier nuclei). Therefore, $Q = +0.782\,$MeV and $-1.804\,$MeV for processes (\ref{eqn:ntop}) and (\ref{eqn:pton}), respectively. In particular, if an antineutrino does not have enough energy to turn a proton into a neutron plus a positron, then that reaction will not proceed, and therefore, the $\bar\nu_e$ on $p$ cross-section is zero for $E_\nu < 1.804\,$MeV. For $\nu_e$ on $n$, however, there is no threshold, and therefore the cross-section is always positive definite. The important thing to note about Eq. (\ref{eqn:sigma}) is the dependence on $E_\nu^2$, implying that higher energy neutrinos would have a stronger effect on the electron fraction. Consequently, the high energy electron neutrino tail which develops in both two- and three-flavor simulations in the bipolar spectral swap case could have a non-negligible effect on the electron fraction.

Using Eqs. (\ref{eqn:lambdanue}) and (\ref{eqn:sigma}), and taking the far-field limit ($r \gg R_\nu$), one can write
\begin{equation}\label{eqn:lambdaav}
\begin{split}
\lambda_{\nu_en}(r)	\approx \frac{\mathcal{C}}{2\pi r^2} \left( \langle E_{\nu_e}^2(r)\rangle + 2\,Q\,\langle E_{\nu_e}(r) \rangle + Q^2 \right)\, \mathcal{N}_{\nu_e}(r), 
\end{split}
\end{equation}
where the averages $\langle E_{\nu_e}(r) \rangle$ and $\langle E_{\nu_e}^2(r) \rangle$ have been calculated with respect to the weighting function
\begin{equation}
f'_{\nu_e}(r,E_\nu) \equiv {\sum}_\alpha \frac{L_{\nu,\alpha}}{\langle E_{\nu,\alpha}\rangle} \tilde{f}_{\nu,\alpha}(E_\nu)P_{\alpha e}(r,E_\nu),
\end{equation}
which can be recognized as the effective electron-flavor neutrino distribution function (non-normalized) at a radius $r$, with
\begin{equation}
\mathcal{N}_{\nu_e}(r) \equiv \int\displaylimits_0^\infty \sum_\alpha \frac{L_{\nu,\alpha}}{\langle E_{\nu,\alpha}\rangle} \tilde{f}_{\nu,\alpha}(E_\nu)P_{\alpha e}(r,E_\nu)\, dE_\nu
\end{equation}
being the effective number luminosity of electron-flavor neutrinos at a radius $r$. For instance, the expression for average electron neutrino energy-squared at a radius $r$ can be calculated as
\begin{equation}
\langle E_{\nu_e}^2(r) \rangle = \frac{1}{\mathcal N_{\nu_e}(r)}\int\displaylimits_0^\infty E_\nu^2 \, f'_{\nu_e}(r,E_\nu) \,dE_\nu
\end{equation}

Armed with this, we can calculate the effective electron neutrino number luminosities and average energies and thereby get an idea of whether, or under what circumstances (e.g., outflow speeds), the neutrinos can have an effect on the electron fraction at different radii within the envelope. Tables \ref{table:rateshigh} and \ref{table:rateslow} list the values of quantities $\langle E_{\nu_e} \rangle$, $\langle E_{\nu_e}^2 \rangle$, and $\mathcal{N}_{\nu_e}$, along with the calculated $\lambda_{\nu_en}$ capture rates, for the bipolar spectral swap simulations with high and low luminosities/matter densities, as discussed in Secs. \ref{sec:2fbipolarresult} and \ref{lowlumlowdens}, at radii of $r = 2000$ km and $r = 1200$ km, respectively. In each table, for comparison we also present a second set of values calculated at these radii, but using the unaltered initial neutrino energy spectra, i.e., assuming that no flavor evolution occurred in the interim (taking $P_{\alpha e}(r,E_\nu) = \delta_{\alpha e}$). The choices of radii were based on the points at which collective flavor oscillations more-or-less ended in the respective simulations.

\begin{table}[ht]
\begin{ruledtabular}
\def\arraystretch{1.2}
\begin{tabular}{ | c | c | c | }
Parameter & No oscillations & With oscillations \\
\colrule
	$\langle E_{\nu_e}(r) \rangle$ & $11$ MeV & $16.6$ MeV \\
   	$\langle E_{\nu_e}^2(r) \rangle$ & $145.6$ MeV$^2$ & $387.6$ MeV$^2$ \\
	$\mathcal{N}_{\nu_e}(r)$ & $8.5\times10^{57}$ s$^{-1}$ & $8.5\times10^{57}$ s$^{-1}$ \\
	$\lambda_{\nu_en}(r)$ & $0.51$ s$^{-1}$ & $1.3$ s$^{-1}$
\end{tabular}
\caption{Table showing values of average energy, average energy-squared, and effective number luminosity in the electron-flavor, along with the calculated charged-current capture rate $\lambda_{\nu_en}$, in the two-flavor bipolar spectral swap simulation with high luminosity and matter density (Table~\ref{table:params}). The numbers presented above are evaluated at a simulation radius of $r = 2000$ km. The numbers in the second column are calculated assuming no neutrino flavor evolution occurs in the interim, whereas those in the third column reflect the changes due to the flavor evolution, corresponding to the results of that simulation.}
\label{table:rateshigh}
\end{ruledtabular}
\end{table}

\begin{table}[ht]
\begin{ruledtabular}
\def\arraystretch{1.2}
\begin{tabular}{|c|c|c|}
Parameter & No oscillations & With oscillations \\
\colrule
	$\langle E_{\nu_e}(r) \rangle$ & $11$ MeV & $15.5$ MeV \\
   	$\langle E_{\nu_e}^2(r) \rangle$ & $145.6$ MeV$^2$ & $354.8$ MeV$^2$ \\
	$\mathcal{N}_{\nu_e}(r)$ & $8.5\times10^{56}$ s$^{-1}$ & $7.6\times10^{56}$ s$^{-1}$ \\
	$\lambda_{\nu_en}(r)$ & $0.14$ s$^{-1}$ & $0.3$ s$^{-1}$
\end{tabular}
\caption{Same as table~\ref{table:rateshigh}, but for the bipolar spectral swap simulation with low luminosity and matter density (Sec.~\ref{lowlumlowdens}), at a radius $r = 1200$ km.}
\label{table:rateslow}
\end{ruledtabular}
\end{table}

For comparison, the rate calculations for the simulation that used the DD2 equation of state luminosities and average energies from Ref.~\cite{2016PhRvD..93d4019F} are presented in Table~\ref{table:ratesdd2}. Even though this simulation exhibited qualitatively similar flavor transformation features, including a bipolar spectral swap at low energies, and a high-energy electron neutrino tail, the lower initial $\nu_x$ luminosity and the relatively weaker energy hierarchy between $\nu_e$ and $\nu_x$ rendered the resulting high-energy electron neutrino tail less potent, both in terms of energy and number. The $\sim 40\%$ change in $\lambda_{\nu_en}$, although less drastic than for the case presented in Table~\ref{table:rateshigh}, can nevertheless be significant for determining $Y_e$ and the corresponding effects on nucleosynthesis.

\begin{table}[ht]
\begin{ruledtabular}
\def\arraystretch{1.2}
\begin{tabular}{ | c | c | c | }
Parameter & No oscillations & With oscillations \\
\colrule
	$\langle E_{\nu_e}(r) \rangle$ & $11.9$ MeV & $14.4$ MeV \\
   	$\langle E_{\nu_e}^2(r) \rangle$ & $169.3$ MeV$^2$ & $266.3$ MeV$^2$ \\
	$\mathcal{N}_{\nu_e}(r)$ & $8.4\times10^{57}$ s$^{-1}$ & $7.6\times10^{57}$ s$^{-1}$ \\
	$\lambda_{\nu_en}(r)$ & $0.59$ s$^{-1}$ & $0.82$ s$^{-1}$
\end{tabular}
\caption{Same as table~\ref{table:rateshigh}, but for the two-flavor bipolar spectral swap simulation run using the initial luminosities and spectra adopted from the DD2 equation of state simulation in Ref.~\cite{2016PhRvD..93d4019F}. The numbers presented above are evaluated at a simulation radius of $r = 2000$ km.}
\label{table:ratesdd2}
\end{ruledtabular}
\end{table}

Table ~\ref{table:rates3f} shows the corresponding rates for a three-flavor calculation exhibiting the bipolar spectral swap (Sec.~\ref{sec:3fbipolarresult}). In this case, the enhancement of the $\nu_e$ capture rate stemming from flavor transformations is less drastic as compared to the result in the corresponding two-flavor case (Table~\ref{table:rateshigh}). However, the effect of the high-energy tail still stands out: despite the effective number luminosity $\mathcal{N}_{\nu_e}$ of electron neutrinos dropping by almost a factor of two from the initial luminosity, the total rate is nevertheless enhanced by about 30\%. This can be attributed to the presence of the high-energy tail and the strong energy dependence of the weak capture cross sections.

\begin{table}[ht]
\begin{ruledtabular}
\def\arraystretch{1.2}
\begin{tabular}{ | c | c | c | }
Parameter & No oscillations & With oscillations \\
\colrule
	$\langle E_{\nu_e}(r) \rangle$ & $11$ MeV & $16.5$ MeV \\
   	$\langle E_{\nu_e}^2(r) \rangle$ & $145.6$ MeV$^2$ & $377.3$ MeV$^2$ \\
	$\mathcal{N}_{\nu_e}(r)$ & $8.5\times10^{57}$ s$^{-1}$ & $4.4\times10^{57}$ s$^{-1}$ \\
	$\lambda_{\nu_en}(r)$ & $0.51$ s$^{-1}$ & $0.65$ s$^{-1}$
\end{tabular}
\caption{Same as table~\ref{table:rateshigh}, but for the three-flavor bipolar spectral swap simulation (Sec.~\ref{sec:3fbipolarresult}), at a radius $r = 2000$ km.}
\label{table:rates3f}
\end{ruledtabular}
\end{table}

Lastly, the capture rate calculations corresponding to the MNR simulation in Sec.~\ref{sec:MNR} are presented, for the 2-flavor and 3-flavor cases respectively, in tables ~\ref{table:ratesMNR} and ~\ref{table:rates3fMNR}. In the 2-flavor case, neutrino flavor evolution boosts the rate $\lambda_{\nu_en}$ not only through the high-energy electron neutrinos in the tail, but also through a net increase in the total number luminosity of electron neutrinos. Nevertheless, because of the weaker hierarchy in average neutrino energies in this case, the effect is still less drastic compared to that shown in table~\ref{table:rateshigh}. The 3-flavor MNR case closely mimics the 2-flavor case in terms of the effect flavor transformations have on the rate $\lambda_{\nu_en}$. A net increase in the average energy of the electron neutrinos boost the neutrino capture rates substantially. The net number of electron neutrinos does not decrease significantly (as might be inferred from figure~\ref{fig:3fMNR}) due to initial Muon and Tau neutrinos transforming resonantly into the electron flavor state. For the 3-flavor case we chose to calculate these rates at a radius of $r=1000\,\text{km}$ because, as we can see from figure~\ref{fig:3fMNR}, the collective flavor evolution appears to stabilize temporarily at this radius, and then resume again past a radius of approximately $r\approx 2000\,\text{km}$. The fact that these rates are approximately a factor of four greater than the rates presented in table~\ref{table:ratesMNR} comes merely from the fact that these rates were calculated closer in towards the merger remnant.

\begin{table}[ht]
\begin{ruledtabular}
\def\arraystretch{1.2}
\begin{tabular}{ | c | c | c | }
Parameter & No oscillations & With oscillations \\
\colrule
	$\langle E_{\nu_e}(r) \rangle$ & $10.6$ MeV & $13.2$ MeV \\
   	$\langle E_{\nu_e}^2(r) \rangle$ & $135.2$ MeV$^2$ & $232.4$ MeV$^2$ \\
	$\mathcal{N}_{\nu_e}(r)$ & $8.8\times10^{56}$ s$^{-1}$ & $1\times10^{57}$ s$^{-1}$ \\
	$\lambda_{\nu_en}(r)$ & $0.05$ s$^{-1}$ & $0.09$ s$^{-1}$ \\
\end{tabular}
\caption{Same as table~\ref{table:rateshigh}, but for the calculation that exhibits the matter-neutrino resonance (Table~\ref{table:MNRparams}), at a radius $r = 2000$ km.}
\label{table:ratesMNR}
\end{ruledtabular}
\end{table}

\begin{table}[ht]
	\begin{ruledtabular}
		\def\arraystretch{1.2}
		\begin{tabular}{ | c | c | c | }
			Parameter & No oscillations & With oscillations \\
			\colrule
			$\langle E_{\nu_e}(r) \rangle$ & $10.6$ MeV & $14.0$ MeV \\
			$\langle E_{\nu_e}^2(r) \rangle$ & $135.2$ MeV$^2$ & $241.0$ MeV$^2$ \\
			$\mathcal{N}_{\nu_e}(r)$ & $8.8\times10^{56}$ s$^{-1}$ & $8.2\times10^{56}$ s$^{-1}$ \\
			$\lambda_{\nu_en}(r)$ & $0.20$ s$^{-1}$ & $0.32$ s$^{-1}$ \\
		\end{tabular}
		\caption{Same as table~\ref{table:ratesMNR}, but for a 3-flavor calculation at a radius $r = 1000$ km.}
		\label{table:rates3fMNR}
	\end{ruledtabular}
\end{table}

In both the high and low luminosity/matter density simulations (Tables~\ref{table:rateshigh} and \ref{table:rateslow}), we see that the rates $\lambda_{\nu_en}$ calculated using our observed flavor transformation are greater than those with no flavor transformation, by factors of two to three. This is not surprising, considering that in both simulations, a strong high-energy electron neutrino tail develops, which skews the average energy and energy-squared towards higher values. To determine whether these neutrinos actually have any purchase on the electron fraction, one must know the local outflow rate of the material in the envelope. Conversely, we can use our neutrino capture rate to estimate what the local outflow velocity $v_\text{out}$ would have to be at any radius along our trajectory in order to effectively decouple the neutrinos, so that the neutron excess can be preserved to facilitate the $r$-process. Neutrino decoupling necessarily requires
\begin{equation}
\frac{v_\text{out}}{r} \gg \lambda_{\nu_en}.
\end{equation}

This implies that, to completely decouple the neutron excess from the neutrinos, the outflow velocities would have to be much greater than $v_\text{out} \approx 2600$ km/s and $v_\text{out} \approx 360$ km/s, for the rates presented in the right-hand columns of tables \ref{table:rateshigh} and \ref{table:rateslow}, respectively. Therefore, as long as the outflow velocities are comparable to these numbers or smaller, the neutrinos would likely be coupled to the electron fraction in the matter, rendering neutrino flavor evolution potentially important in determining the $r$-process production feasibility for the wind-like ejecta outside the neutrino disk plane.

We can see from Eq. (\ref{eqn:YeEvol}) that an increase in the cross section for neutrino capture, and therefore and increase in the rate $\lambda_{\nu_e n}$, would tend to raise the electron fraction $Y_e$. Even though this rate appears in both the positive and negative parts of the differential equation, the negative part is multiplied by $Y_e$ itself which must be less than 1. The net effect of increasing this rate, then, would be to increase $Y_e$ towards 1. This makes sense as this rate is a rate for a reaction which destroys neutrons and creates protons. If the high energy electron neutrino tail would cause the $Y_e$ to rise above the level that current simulations without neutrino flavor evolution account for, then this would generally hurt the efficiency of the $r$-process. For a robust $r$-process, there must be a sufficiently large ratio of neutrons to seed nuclei, usually implying the necessity of a low electron fraction.

\section{Conclusions}\label{sec:conclusions}

We have investigated flavor transformation phenomena for polar-axis directed neutrinos streaming out from a BNS merger neutrino disk. In cases where the total number luminosity of neutrinos is higher than antineutrinos, we have seen that neutrino flavor transformations in a BNS merger neutrino-driven wind may give rise to a bipolar spectral swap at low energies, along with a high energy electron neutrino tail, in the normal mass hierarchy. Such a scenario (neutrino number dominated) can arise in merger simulations with the DD2 neutron star equation of state. The bipolar spectral swaps found in our results are qualitatively similar to those obtained from flavor transformation simulations in supernova environments, demonstrating the robustness of the mechanism underlying the swap to the geometric differences between the two cases. In our calculations, this phenomenon was observed in simulations with varying luminosities and matter densities, as long as the total number luminosity of electron neutrinos was higher than that of electron antineutrinos. In fact, bipolar oscillations in BNS merger environments were also found in Ref.~\cite{2016arXiv160705938F} for anti-neutrino dominated spectra on certain trajectories. However, those calculations used the inverted mass hierarchy. For the case with a higher electron antineutrino number luminosity, we were able to qualitatively reproduce the matter-neutrino-resonance (MNR) that was observed in previous studies of the binary neutron star merger environment. In both cases, the high energy tail which develops in the electron neutrino spectrum, with the absence of an analogous phenomenon in the antineutrino sector, serves to enhance the charged-current neutrino capture rate on neutrons. In the absence of rapid matter outflows, this increase in the capture rate could lead to reduction in the neutron fraction, and thereby a less efficient $r$-process than would be expected if neutrino flavor evolution were not taken into account. 

It is intriguing that aspects of the hot, neutron matter equation of state which determine the emergent neutrino energy spectra and fluxes, also may qualitatively influence the nature and outcome of collective neutrino oscillations and consequently, the outflow composition in some cases.

\begin{acknowledgements}
We would like to thank J.~Carlson, J.F.~Cherry, V.~Cirigliano, L.~Johns, C.~Kishimoto, J.T.~Li, S.~Tawa, and A.~Vlasenko, for valuable conversations. This work was supported in part by NSF Grants PHY-1307372 and PHY-1614864 at UCSD. We also acknowledge a grant from the University of California Office of the President.
\end{acknowledgements}

\bibliography{allref}

\end{document}